\documentclass{article}
\usepackage{hyperref}
\usepackage{enumerate}
\usepackage{color, soul}
 \usepackage{metalogo} %for XeLaTeX logo
 \usepackage{algpseudocodex}
\usepackage[numbers]{natbib}
\usepackage{amsmath,amssymb,amsfonts}
\usepackage{graphicx}
\usepackage{bbm}
\usepackage{mathrsfs}%
\usepackage{booktabs}
\begin{document}
\title{A probabilistic diagnostic for Laplace approximations: Introduction and experimentation}
\maketitle
\author{Shaun McDonald (Corresponding Author)}\footnotemark[1]
\author{David Campbell}\footnotemark[2]

\footnotetext[1]{Department of Statistics and Actuarial Science, Simon Fraser Univesity, 8888 University Drive, Burnaby, V5A 1S6, British Columbia, Canada. drshaunmcdonald at gmail dot com}
\footnotetext[2]{School of Mathematics and Statistics, Carleton University, 1125 Colonel By Drive, Ottawa, K1S 5B6, Ontario, Canada. davecampbell at math dot carleton dot ca}

\begin{abstract}
Many models require integrals of high-dimensional functions: for instance, to obtain marginal likelihoods. Such integrals may be intractable, or too expensive to compute numerically. Instead, we can use the \textit{Laplace approximation} (LA). The LA is exact if the function is proportional to a normal density; its effectiveness therefore depends on the function's true shape. Here, we propose the use of the \textit{probabilistic numerical} framework to develop a diagnostic for the LA and its underlying shape assumptions, modelling the function and its integral as a Gaussian process and devising a ``test'' by conditioning on a finite number of function values. The test is decidedly non-asymptotic and is not intended as a full substitute for numerical integration - rather, it is simply intended to test the feasibility of the assumptions underpinning the LA with as minimal computation. We discuss approaches to optimize and design the test, apply it to known sample functions, and highlight the challenges of high dimensions.
\end{abstract}
%
%\begin{frenchabstract}
%De nombreux mod\`{e}les n\'{e}cessitent des int\'{e}grales de fonctions de grande dimension: par exemple, pour obtenir des vraisemblances marginales. De telles int\'{e}grales peuvent \^{e}tre insolubles ou trop coûteuses \`{a} calculer num\'{e}riquement. Au lieu de cela, nous pouvons utiliser l'\textit{Approximation de Laplace} (AL). Le AL est exact si la fonction est proportionnelle \`{a} une densit\'{e} normale; son efficacit\'{e} d\'{e}pend donc de la v\'{e}ritable forme de la fonction. Ici, nous proposons l'utilisation du cadre \textit{num\'{e}rique probabiliste} pour d\'{e}velopper un diagnostic pour le AL et ses hypoth\`{e}ses de forme sous-jacentes, en mod\'{e}lisant la fonction et son int\'{e}grale comme un processus gaussien et en concevant un ``test'' en conditionnant sur un nombre fini de valeurs de fonction. Le test est r\'{e}solument non asymptotique et n'est pas destin\'{e} \`{a} remplacer compl\`{e}tement l'int\'{e}gration num\'{e}rique. Il vise plut\^{o}t simplement \`{a} tester la faisabilit\'{e} des hypoth\`{e}ses qui sous-tendent l'AL avec un minimum de calculs. Nous discutons des approches pour optimiser et concevoir le test, l'appliquons \`{a} des exemples de fonctions connus et mettons en \'{e}vidence les d\'{e}fis des grandes dimensions.
%\end{frenchabstract}

%\begin{keywords}
%\item Probabilistic numerics
%\item Laplace approximation
%\item State-space modelling
%\item Bayesian quadrature
%\end{keywords}

\section{Introduction} \label{sec:laplace_intro}
Many statistical models assume the existence of ``unseen'' variables which influence the actual observed data, but are distinct from the model parameters that are of interest for inference. One such model is the \textit{state-space model} (SSM), which has become a staple of ecological modelling \citep[e.g.][and references therein]{Aeberhard2018} and will serve as a motivating example throughout this manuscript. Briefly, the SSM assumes that (possibly vector-valued) data $\mathbf{y}_t$ are observed at discrete time steps $t \in \{1, \dots, T\}$. At a given time $t$, the distribution of $\mathbf{y}_t$ depends on an unobserved or ``hidden'' state $\mathbf{x}_t \in \mathbb{R}^q$ (typically the dimensionality of $\mathbf{x}_t$ is the same for all $t$, but it may differ from the dimensionality of the $\mathbf{y}_t$'s). In turn, the distribution of $\mathbf{x}_t$ depends on the previous hidden state, $\mathbf{x}_{t-1}$. The reader may recognize this as the structure of a \textit{hidden Markov model (HMM)}, although that term is typically used when the domain of the hidden states is discrete \citep[e.g][]{Chopin2020}. Here, they are assumed to be continuous and possibly multivariate.

In mathematical terms, the SSM is characterized by the joint likelihood\footnotemark
\footnotetext{There are several possible formulations for the distribution of the first hidden state. Some literature assumes it to depend on an ``initial state'' which is given its own prior in turn \citep[e.g.][]{Murray2015} or simply point estimated \citep[e.g.][]{Skaug2006}. The latter is essentially equivalent to specifying an ``unconditional'' distribution for the first state, another common approach \citep[e.g.][]{Chopin2020, Koyama2010}. Some authors omit the first term term entirely, thereby implicitly assigning the first state an ``improper uniform prior'' \citep[e.g.][which is the formulation used in Section \ref{sec:cod}]{Nielsen2014}. The general model form given in equation~\eqref{eq:SSM_joint_lik} will suffice for the purposes of this manuscript.}
\begin{align}
p_{x,y}\left(\mathbf{x}, \mathbf{y} \mid \mathbf{\theta}\right) = p\left(\mathbf{x}_1 \mid \mathbf{\mathbf{\theta}}\right)\left[\prod_{t=2}^T p\left(\mathbf{x}_t \mid \mathbf{x}_{t-1}, \mathbf{\theta}\right)\right]\left[\prod_{t=1}^T p\left(\mathbf{y}_t \mid \mathbf{x}_t, \mathbf{\theta}\right)\right], \label{eq:SSM_joint_lik}
\end{align}
where $\mathbf{x} = \left(\mathbf{x}_1, \dots, \mathbf{x}_T\right)$ is a vector of dimension $d = qT$ concatenating the hidden states, $\mathbf{y}$ is defined analogously, and $\mathbf{\theta}$ is a vector of model parameters. These parameters are conceptually different from the hidden states even though both are unobserved: $\mathbf{\theta}$ represents the \textit{fixed effects} of the model, whereas $\mathbf{x}$ represents \textit{random effects}\footnotemark.
\footnotetext{Of course, in a Bayesian setting, both model components are given priors and essentially treated in the same way. In that case, the difference between them is more of a ``philosphical'' matter.}

There are a variety of methods for both frequentist and Bayesian inference with SSM's \citep[e.g.][and references therein]{DeValpine2002, Skaug2006}. In the frequentist framework, one typically wishes to estimate $\mathbf{\theta}$ by maximizing the marginal likelihood of the data,
\begin{align}
p_y\left(\mathbf{y} \mid \mathbf{\theta}\right) = \int_{\mathbb{R}^d} p_{x,y}\left(\mathbf{x}, \mathbf{y} \mid \mathbf{\theta}\right) \mathrm{d}\mathbf{x}. \label{eq:SSM_marg_lik}
\end{align}
Unfortunately, the necessary integral over the hidden states is $d$-dimensional, and as such the marginal likelihood cannot realistically be computed --- much less optimized --- in most cases. Instead, frequentist inference methods for SSM's typically rely on approximations of various types to obtain a suitable estimate of $\mathbf{\theta}$. Examples include methods based on particle filtering, as described by \citet{Kantas2015}. Another common --- and less computationally demanding \citep[e.g.][]{Aeberhard2018} --- approach is use of the \textit{Laplace approximation} (LA). The Laplace approximation of the marginal likelihood is reasonably easy to compute and optimize as a function of $\mathbf{\theta}$, but it is based on certain assumptions about the shape of the joint likelihood as a function of $\mathbf{x}$: namely, that it is well approximated by a $d$-dimensional Gaussian density. If this assumption is not satisfied, the LA may not be suitable, and different methods for SSM inference may need to be invoked.

The example of the SSM provides motivation for the broader goal of this manuscript, which is to develop a diagnostic tool to check the assumptions underpinning the LA. In particular, our interest is in assessing whether or not a given function is ``close enough'' to the Gaussian shape to justify using the Laplace approximation of its integral. In making this assessment, we strive for a ``middle ground'' of computational effort: the diagnostic will naturally be more complex than the LA itself, but much less expensive than a full-fledged numerical estimate of the integral. Expanding on the work of \citet{Zhou2017}, here we describe such a diagnostic tool based on the machinery of \textit{probabilistic numerics}, a burgeoning field which exploits probability theory to tackle numerical problems. The tool is an application of the probabilistic numerical technique of \textit{Bayesian quadrature} (BQ), which allows for both estimation and inference of unknown integrals. Unlike ``conventional'' BQ, however, the actual integral value is of secondary importance, as the tool is primarily intended to capture as much information as possible about the \textit{shape of the integrand}. In keeping with the aforementioned objective of ``medium effort'', the tool is also decidedly non-asymptotic: it is meant to deliver as much information as possible with a modest amount of computation, without consideration of any type of limiting behaviour. The goal is a fast, informal method that can be readily deployed to determine if additional modelling efforts are needed beyond the LA.

The remainder of the manuscript proceeds as follows. Section \ref{sec:framework} defines the LA and establishes the notation used throughout this manuscript, while Section \ref{sec:pn} provides more detail about the workings of probabilistic numerics and BQ in particular. Sections \ref{sec:design}--\ref{sec:cal} provide technical details about the design of our diagnostic tool, and Section \ref{sec:banana} shows a low-dimensional application. Section \ref{sec:high}  is focused on challenges, applications, and discussion of the diagnostic in high-dimensional settings.

\section{Framework and notation} \label{sec:framework}
Consider a positive function $f: \mathbb{R}^d \to \mathbb{R}_{>0}$ and its integral $F = \int_{\mathbb{R}^d} f(\mathbf{x}) \mathrm{d}\mathbf{x}$.
%In practical applications \texttt{[citation needed]}, typically $f$ and $F$ are actually functions with additional arguments comprising data $\mathbf{y}$ and structural parameters $\mathbf{\theta}$, with $x \in \mathbb{R}^d$ a vector of nuisance parameters to be marginalized. For instance, $f$ may be a joint probability density for $\left(\mathbf{y}, x \mid \mathbf{\theta}\right)$, in which case $F$ would be the marginal distribution of $\mathbf{y} \mid \mathbf{\theta}$ after integrating over $x$. To reflect this common setting, \citet{Zhou2017} called $f$ and $F$ the \textit{full} and \textit{target} functions, respectively. For the present discussion, non-marginalized arguments $\mathbf{y}$ and $\mathbf{\theta}$ are not relevant, so any dependence on them is omitted and $f$ and $F$ are simply called the \textit{true} function and integral, respectively.
More rigorous treatments of the Laplace approximation are available in, for instance, \citet{de1981asymptotic} and \citet{barndorff1989asymptotic}, but for this exposition it suffices to assume that all second-order partial derivatives of $f$ exist and are continuous, and that $f$ attains a maximum at some point $\hat{\mathbf{x}} \in \mathbb{R}^d$. To reflect the common use case where $f$ is a density or likelihood, $\hat{\mathbf{x}}$ is called a \textit{mode}. Let $H$ be the Hessian of $\log{f}$ at $\hat{\mathbf{x}}$ and suppose that it is negative definite.
Taking a second-order Taylor expansion of $\log{f}$ about $\hat{\mathbf{x}}$ gives the approximation
\begin{align}
\log{f}(\mathbf{x}) \approx \log{f\left(\hat{\mathbf{x}}\right)} + \frac{1}{2}\left(\mathbf{x} - \hat{\mathbf{x}}\right)^T \mathbf{H}\left(\mathbf{x} - \hat{\mathbf{x}}\right), \label{eq:taylor}
\end{align}
since all first-order partial derivatives of $\log{f}$ are equal to zero at the mode. Exponentiating approximation~\eqref{eq:taylor} gives an approximation for $f$ in the form of (up to normalizing constants) a Gaussian density centered at $\hat{\mathbf{x}}$ with covariance matrix $-\mathbf{H}^{-1}$. In turn, integrating this exponentiated function (hereafter called the \textit{Gaussian approximation to $f$}) produces the \textit{Laplace approximation} to $F$:
\begin{align}
F \approx L\left(f\right) &:= f\left(\hat{\mathbf{x}}\right)\int_{\mathbb{R}^d} \exp{\left[\frac{1}{2}\left(\mathbf{x} - \hat{\mathbf{x}}\right)^T \mathbf{H}\left(\mathbf{x} - \hat{\mathbf{x}}\right)\right]}\mathrm{d}\mathbf{x} \nonumber \\
&= f\left(\hat{\mathbf{x}}\right)\sqrt{\left(2\pi\right)^d \det{\left(-\mathbf{H}^{-1}\right)}}. \label{eq:lap_approx}
\end{align}
The LA has a long history of use in statistics \citep[e.g.][]{Lindley1961, Tierney1986}. It is exact (or ``true'') if the integrand $f$ is itself proportional to a Gaussian density. There are other function shapes for which this may be the case, but such instances may be thought of as ``coincidence''. Certainly, the derivation of the LA via approximation~\eqref{eq:taylor} is based on an assumption of approximately Gaussian shape (insofar as it assumes that the second-order Taylor series is a reasonable approximation to $\log f$), and as noted in Section \ref{sec:laplace_intro}, this assumption is our main interest.

Before proceeding to further details about the construction of the diagnostic tool, it is worthwhile to connect these concepts to the SSM example described in Section \ref{sec:laplace_intro}. For given observations $\mathbf{y}$ and parameter values $\mathbf{\theta}$, the joint likelihood $p_{xy}\left(\cdot, \mathbf{y} \mid \mathbf{\theta}\right)$ takes the role of the integrand, viewed as a function of the hidden states $\mathbf{x} \in \mathbb{R}^d$. In turn, one can see from equation~\eqref{eq:SSM_marg_lik} that the marginal likelihood $p_y\left(\mathbf{y} \mid \mathbf{\theta}\right)$ takes the role of the integral over $\mathbb{R}^d$ to be approximated by $L\left(p_{xy}\right)$. Note, however, that this approximation is itself a function of $\mathbf{y}$ and $\mathbf{\theta}$, as both
\begin{align}
\hat{\mathbf{x}} = \underset{\mathbf{x}}{\mathrm{argmax}}\ p_{xy}\left(\mathbf{y}, \mathbf{x} \mid \mathbf{\theta}\right) \quad\mathrm{and}\quad \mathbf{H} = \frac{\partial^2 \log p_{xy}}{\partial \mathbf{x}^2} \bigg\rvert_{\left(\mathbf{y}, \hat{\mathbf{x}}, \mathbf{\theta}\right)}\nonumber
\end{align}
may depend on these quantities. Indeed, one of the most common ways to ``fit an SSM'' in the frequentist sense is to maximize $L\left(p_{xy}\right)$ with respect to $\mathbf{\theta}$ (given observed $\mathbf{y}$), typically using standard numerical algorithms. Fitting the model in this way becomes a matter of \textit{nested} optimization, since in each iteration $\hat{\mathbf{x}} = \hat{\mathbf{x}}(\mathbf{\theta}, \mathbf{y})$ must be (numerically) calculated for the current $\mathbf{\theta}$-value \citep[see][for instance]{Kristensen2016}.

Implicit in the use of such methods for SSM's is the assumption that the LA is reasonably accurate given $\mathbf{y}$ and for each $\mathbf{\theta}$-value calculated during the optimization steps. If the shape of $p_{xy}$ with respect to $\mathbf{x}$ is not ``sufficiently Gaussian'' at a given iteration, then the ultimate estimate of $\mathbf{\theta}$ may not be close to the actual MLE for the marginal likelihood. Therefore, it would be desirable to check the validity of the LA at each step, using the diagnostic tool detailed below.

\section{Probabilistic numerics and Bayesian quadrature} \label{sec:pn}
Broadly speaking, probabilistic numerics is the use of probability theory, from a somewhat Bayesian perspective, to simultaneously perform estimation and uncertainty quantification in standard numerical problems \citep{Hennig2015}. For instance, \citet{Chkrebtii2016} developed a probabilistic solver for differential equations. For a given equation, they jointly modelled the function and its derivatives with a Gaussian process prior, then sequentially conditioned on evaluations of the true derivative to conduct posterior inference on the entire solution.

The approach briefly described above --- using Gaussian process priors and finitely many function evaluations to obtain posteriors for the functions and quantities of interest --- is at the core of many probabilistic numerical methods. In particular, it is the standard framework with which \textit{Bayesian quadrature} (BQ) is usually conducted \citep[see][and references therein]{Briol2019, Cockayne2019}. As the name suggests, BQ is a probabilistic analogue to standard numerical integration that uses a combination of prior belief and gathered information about a function. The remainder of this section, in which the diagnostic for the LA is developed, will also serve as an explanation of the mathematical machinery underpinning BQ.

Literature on BQ commonly assumes that the integral of interest is with respect to a probability (i.e.\ finite) measure $G$ on the domain \citep[e.g][]{Briol2019}, and a standard choice for $\mathbb{R}^d$ is a $d$-dimensional Gaussian measure \citep{OHagan1991, Karvonen2018}. Accordingly, we use an ``importance weighting trick'' \citep{Kennedy1998, Rasmussen2003, Osborne2010} to re-express the integral of interest. Recalling the notation of Section \ref{sec:framework}, the integral of $f$ over $\mathbb{R}^d$ is
\begin{align}
F = \int_{\mathbb{R}^d} f(\mathbf{x}) \mathrm{d}\mathbf{x}  = \int_{\mathbb{R}^d} r(\mathbf{x}) g(\mathbf{x}) \mathrm{d}\mathbf{x} = \int_{\mathbb{R}^d} r(\mathbf{x}) \mathrm{d}G\left(\mathbf{x}\right), \label{eq:int_weight}
\end{align}
where $r :=  f/g$ and $g$ is the density of the aforementioned Gaussian measure $G$, the parameters of which will be discussed later. It is this ``re-weighted'' function $r$ that is modelled with a Gaussian process prior \citep{Kennedy1998}. The mean function of the GP prior, $m_0^x$, is taken to be the (similarly re-weighted) Gaussian approximation of $f$ underpinning approximation~\eqref{eq:taylor} and equation~\eqref{eq:lap_approx}:
\begin{align}
m^x_0(\mathbf{x}) := \frac{f\left(\hat{\mathbf{x}}\right) \exp{\left[\frac{1}{2}\left(\mathbf{x} - \hat{\mathbf{x}}\right)^T H\left(\mathbf{x} - \hat{\mathbf{x}}\right)\right]}}{g(\mathbf{x})}, x \in \mathbb{R}^d. \label{eq:prior_mean}
\end{align}
The covariance operator for the GP is a (positive-definite) kernel $C_0^x$ on $\mathbb{R}^d \times \mathbb{R}^d$, defined in Section \ref{sec:covar}. Because integration is a linear projection, such a prior on $g$ induces a univariate normal prior on $F$ with mean $m_0 := \int_{\mathbb{R}^d} m_0^x(\mathbf{x})\mathrm{d}G\left(\mathbf{x}\right) = L(f)$ and variance  $C_0 := \int_{\mathbb{R}^d}\int_{\mathbb{R}^d}C_0^x(\mathbf{x},\mathbf{z})\mathrm{d}G\left(\mathbf{x}\right)\mathrm{d}G\left(\mathbf{z}\right)$ \citep[e.g.][]{Rasmussen2003, Hennig2015}.

In what follows, let $\mathbf{S} = \left(\mathbf{s}_1, \dots, \mathbf{s}_n\right)^T \in \mathbb{R}^{n \times d}$ be a row-wise concatenation of $n$ (transposed) vectors in $\mathbb{R}^d$ (we will sometimes call it a ``\textit{grid}'' of $n$ ``points'' in $\mathbb{R}^d$). Then, for instance, the notation $\mathbf{r}\left(\mathbf{S}\right)$ will refer to the column vector $\left(r\left(\mathbf{s}_1\right), \dots, r\left(\mathbf{s}_n\right)\right)^T \in \mathbb{R}^n$, and $\mathbf{C}_0^x\left(\mathbf{S}, \mathbf{S}\right)$ will denote the $n\times n$ matrix with $(i, j)^\mathrm{th}$ entry $C_0^x\left(\mathbf{s}_i, \mathbf{s}_j\right)$. Using standard GP identities \citep[e.g.][]{Rasmussen2006}, one may use true function values at the \textit{interrogation points} $\mathbf{S}$ to obtain a posterior distribution for $g$ (with another slight abuse of notation):
\begin{align}
r \mid \mathbf{r}(\mathbf{S}) &\sim \mathcal{GP}\left(m_1^x, C^x_1\right), \\
m_1^x(\mathbf{x}) &= m_0^x(\mathbf{x}) + \mathbf{C}^x_0(\mathbf{x}, \mathbf{S})^T\left[\mathbf{C}_0^x(\mathbf{S}, \mathbf{S})\right]^{-1}\left(\mathbf{r}(\mathbf{S}) - \mathbf{m}_0^x(\mathbf{S})\right), \label{eq:mean} \\
C_1^x(\mathbf{x},\mathbf{z}) &= C_0^x(\mathbf{x},\mathbf{z}) - \mathbf{C}_0^x(\mathbf{x}, \mathbf{S})^T \left[\mathbf{C}^x_0(\mathbf{S}, \mathbf{S})\right]^{-1}\mathbf{C}_0^x(\mathbf{z}, \mathbf{S}).
\end{align}
In turn, the posterior distribution on the integral $F$ is \citep[e.g.][or, indeed, virtually any BQ paper]{Briol2019}
\begin{align}
F \mid \mathbf{r}(\mathbf{S}) &\sim \mathcal{N}\left(m_1, C_1\right), \label{eq:int_posterior} \\
m_1 &= L(f) + \left[\int_{\mathbb{R}^d}\mathbf{C}_0^x(\mathbf{x},\mathbf{S})\mathrm{d}G\left(\mathbf{x}\right)\right]^T\left[\mathbf{C}^x_0(\mathbf{S}, \mathbf{S})\right]^{-1}\left(\mathbf{r}(\mathbf{S}) - \mathbf{m}_0^x(\mathbf{S})\right), \label{eq:int_mean} \\
C_1 &= C_0 - \left[\int_{\mathbb{R}^d}\mathbf{C}_0^x(\mathbf{x}, \mathbf{S})\mathrm{d}G\left(\mathbf{x}\right)\right]^T \left[\mathbf{C}^x_0(\mathbf{S}, \mathbf{S})\right]^{-1}\left[\int_{\mathbb{R}^d}\mathbf{C}_0^x(\mathbf{x}, \mathbf{S})\mathrm{d}G\left(\mathbf{x}\right)\right]; \label{eq:int_variance}
\end{align}
where the integrals in equation~\eqref{eq:int_variance} are row-wise over $\mathbf{S}$: 
\begin{align}
\int_{\mathbb{R}^d}\mathbf{C}_0^x(\mathbf{x}, \mathbf{S})\mathrm{d}G\left(\mathbf{x}\right) = \left(\int_{\mathbb{R}^d}C_0^x\left(\mathbf{x},\mathbf{s}_1\right)\mathrm{d}G\left(\mathbf{x}\right), \dots, \int_{\mathbb{R}^d}C_0^x\left(\mathbf{x},\mathbf{s}_n\right)\mathrm{d}G\left(\mathbf{x}\right)\right)^T. \nonumber
\end{align}
It is useful to think of the posterior means and variances as their prior counterparts modified by the addition or subtraction of some ``correction term''.

The posterior distribution~\eqref{eq:int_posterior} will serve as the diagnostic for the Laplace approximation. Borrowing from the traditional notion of hypothesis testing, one may deem the Laplace approximation (or, perhaps more accurately, the shape assumptions motivating it) acceptable or valid if $L(f)$ falls within the range spanned by the $(0.025, 0.975)$ quantiles of distribution~\eqref{eq:int_posterior}: the 95\% ``confidence interval'' centered at the posterior mean. Conversely, if $L(f)$ is outside of this interval, the Laplace approximation would be deemed inappropriate (``rejection''), and one could proceed to use a more involved method of estimating $F$.
%Traditionally \texttt{[add old BQ citation]}, the goal of BQ is convergence to the true integral: choosing the covariance kernel and interrogation points such that equation~\eqref{eq:int_mean} and equation~\eqref{eq:int_variance} are close to $F$ and $0$, respectively. This is not our main goal in designing the diagnostic, which is intended to be decidedly non-asymptotic: rather, it should be able to effectively facilitate the aforementioned ``hypothesis'' test with as little computational cost as possible, whether or not that results in a good integral estimate.

\section{Design decisions} \label{sec:design}
In general terms, there are three major categories of ``design'' choices one must make in order to conduct BQ, each of which will be explored in the following subsections. First, we must decide where to place interrogation points $\mathbf{S}$; second, a covariance kernel $C^x_0$ must be chosen for the GP prior; and finally, we must specify the measure $G$ against which to integrate. The latter two  involve setting some \textit{hyperparameters} that will govern the behaviour of the Gaussian process; this will be deferred to Section \ref{sec:cal}. %Although it is not commonly discussed in the literature (aside from some brief consideration by \citet{OHagan1991}),

Recall that the diagnostic is intended to quickly --- and somewhat heuristically --- determine whether a given function $f$ is ``sufficiently Gaussian'' to justify the LA for its integral. In particular, it should expend only as much computational effort as is necessary to reliably make this determination, with actual estimation of the integral $F$ being a \textit{secondary} goal. In this respect, its objectives are different from those of ``traditional'' BQ, in which interrogation points may be chosen to minimize the posterior variance of the integral \citep{OHagan1991, Minka2000, Huszar2012} or the entropy of the integrand \citep{Gunter2014}; and hyperparameters may be chosen by some goodness-of-fit criterion \citep{Briol2019, Rasmussen2006} or approximately marginalized \citep{Osborne2012}, with both approaches depending on the ``observations'' $\mathbf{r}\left(\mathbf{S}\right)$. The computational costs arising from such methods would be antithetical to the ``moderately fast'' nature of the diagnostic. Instead, it should be ``one-size-fits-all'' so that it can be quickly applied to any suitable function. Although ``ad hoc'' design choices are made in some BQ papers \citep[e.g][]{Karvonen2018}, the fact remains that the usual goal is to obtain an accurate integral estimate with low uncertainty. Beyond the issue of computation, there is a more fundamental difference between our goals and those of ``traditional'' BQ, or, indeed, the usual principles of inference in a more general sense. Typically, one may wish to maximize the \textit{power} of their inference, ensuring that any true deviation from some null hypothesis will be found with sufficient data. In the present context, this would mean embracing the standard BQ goal of high accuracy and low uncertainty, so that even the smallest deviation from the LA could be rejected if there are enough well-placed interrogation points. However, such a diagnostic would not be very useful in practice. Hearkening back to the SSM example from Section \ref{sec:laplace_intro}, in all but the simplest models it will be known in advance that the joint likelihood is not \textit{exactly} Gaussian, and the LA not exactly met. The pertinent question is whether the joint likelihood is Gaussian \textit{enough}, and a diagnostic that answered this question in the negative for every nonlinear model would be trivial and useless. Thus, the usual aim of high ``power'' is actually \textit{not} desirable here: the diagnostic should be calibrated such that it \textit{fails to reject} any function which is ``close enough'' to Gaussian, in a sense explained below. In this way, the design choices detailed in the following sections target an unconventional notion of ``\textit{good-enough-ness of fit}''.

\subsection{Placement of interrogation points} \label{sec:points}
The selection of interrogation points (or ``nodes'', as they are commonly known in the literature) is the defining feature of any quadrature method. Much has been written about the asymptotic error rates (as the number of points $n \to \infty$) of various quadrature methods, and the ways in which they depend on the dimensionality of the domain $d$ and the smoothness of the integrand \citep[e.g.][]{Kaarnioja2013, Briol2019}. However, none of these considerations are relevant to the development of a quick, one-size-fits-all tool intended to determine if a function is ``Gaussian enough'' for the LA to be reasonable. Thus, the grid of interrogation points must provide as much pertinent information as possible about the \textit{shape} of $f$, and (particularly in high dimensions, as explained in Section \ref{sec:highdim}) how this shape influences the validity of the LA. Importantly, it must do this with as small a grid as possible in order to be ``medium-effort''; in particular, the grid size must grow at a reasonable rate with respect to $d$. One hopes that the goals of the diagnostic can be accomplished with less computation than it takes to conduct a more accurate BQ.

To begin with, let $\mathbf{S}^* = \left(\mathbf{s}_1, \dots, \mathbf{s}^*_n\right)^T \in \mathbb{R}^{n \times d}$ be a grid of ``preliminary'' interrogation points. Ostensibly the preliminary grid should not depend on any properties of the function $f$, but considerations such as dimensionality can certainly inform its construction. We will assume that the grid is a union of \textit{fully symmetric sets}, as considered by \citet{Karvonen2018}. Briefly, this means that if we take an arbitrary vector $\mathbf{s}^*_i$ from the grid, any vector obtained via permutation or sign changes of its coordinates is also in the grid [ibid.]. We also assume that the grid contains multiples of the standard basis vectors of $\mathbb{R}^d$ (i.e.\ points are placed ``along the axes'') and that its centroid is the origin (the origin may be included in the grid, but this is not strictly necessary). No further restrictions will be placed on the preliminary grid, but some type of sparsity is desirable for the computational reasons mentioned above. The sparse grid methods described by \citet{Karvonen2018}, or modifications thereof, are particularly useful to this end.

Now, recalling that $\mathbf{H}$ is negative-definite, consider the eigendecomposition $-\mathbf{H}^{-1} = \mathbf{VDV}^T$ (where $\mathbf{V}$ is orthogonal and $\mathbf{D}$ is diagonal) and let $\mathbf{T} := \mathbf{V}\sqrt{\mathbf{D}}$. %Consider the function $f^*: t \mapsto f\left(Gt + \hat{\mathbf{x}}\right)$, which is centered at the origin and has a Gaussian approximation proportional to a (multivariate) standard Normal density. 
The vectors comprising the actual interrogation grid $\mathbf{S}$ used in the diagnostic will be affine transformations of the preliminary grid vectors:  $\mathbf{s}_i = \mathbf{T}\mathbf{s}^*_i + \hat{\mathbf{x}}$, $i \in \{1, \dots, n\}$. This transformation serves three purposes. The first is a translation so that the centroid of the grid is $\hat{\mathbf{x}}$, the mode of $f$. Since $f$ will be a density or likelihood in most applications, it makes sense for the grid to be oriented around the region of highest density. In contrast, a grid centered at the origin may be ``off-center'' for some integrands, capturing only limited tail behaviour and certainly not enough ``shape information''. The second purpose for the transformation is a rotation, as $\mathbf{T}$ maps standard basis vectors to eigenvectors of $\mathbf{H}$ (which are the same as those of $-\mathbf{H}^{-1}$). Thus, by placing some of the preliminary points along the ``standard axes'' of $\mathbb{R}^d$, we ensure that the corresponding interrogation points are aligned along the directions in which the ``curvature'' of $f$ at the mode is most extreme\footnotemark.\footnotetext{This point can be formalized and made clear with some linear algebra and multivariate calculus. Note that the second directional derivative at the mode is always negative and is maximized (resp.\ minimized) in the direction of the first (resp.\ last) eigenvector of the Hessian.} Because $\mathbf{H}$ completely characterizes the shape of $f$ under the ``null hypothesis'' that it (approximately) satisfies the assumptions of the LA, heuristically it makes sense to say that, \textit{a priori}, one would expect such interrogation points to contain the most pertinent ``shape information''. Finally, the transformation ``stretches'' its inputs in the direction of each eigenvector $\mathbf{V}_i$ by a factor of $\sqrt{D_{ii}}$ ($D_{ii}$ being the eigenvalue associated with $\mathbf{V}_i$). Thus, if $\mathbf{H}$ is such that the Gaussian approximation to $f$ (and, presumably, $f$ itself) has different scales in different directions, the grid will capture this appropriately. In summary, this transformation turns a preliminary grid of the type stipulated above into an interrogation grid that is adapted to the contours of the Gaussian approximation to $f$. In this respect, it can be assumed --- \textit{a priori} or ``under the null hypothesis'' of Gaussian shape --- that the grid so obtained is, in some informal sense, ``optimal'' for obtaining the necessary information about $f$.

%Finally observe that this statement must also be true for $f$ itself since is always positive and its gradient is zero at $\hat{\mathbf{x}}$.}

There is another, perhaps more intuitive interpretation of interrogation grids generated in this way. Let $\mathbf{X}$ be a multivariate normal random variable with density proportional to the Gaussian approximation to $f$, i.e.\ $\mathbf{X} \sim \mathcal{N}(\hat{\mathbf{x}}, -\mathbf{H}^{-1})$. Then the $i^\mathrm{th}$ component of the vector $\mathbf{VX}$ is the $i^\mathrm{th}$ \textit{principal component}, or PC, of $\mathbf{X}$, and has marginal variance equal to $D_{ii}$ \citep{Jolliffe2002}. Thus, the affine transformation of the preliminary grid is centered at the mean of $\mathbf{X}$, aligned with its ``principal axes'', and scaled according to the scales of its PC's. For example, recall that for $i \in \{1, \dots, d\}$, the preliminary grid contains points of the form $\pm m\mathbf{e}_i$, where $m > 0$ and $\mathbf{e}_i$ is the $i^\mathrm{th}$ standard basis vector of $\mathbb{R}^d$. The corresponding interrogation points, $\pm m\sqrt{D_{ii}}\mathbf{V}_i + \hat{\mathbf{x}}$, are ``$m$ standard deviations (of the $i^\mathrm{th}$ PC of $\mathbf{X}$) away from the mode (in the direction of that PC)''.

\subsection{Form of covariance kernel} \label{sec:covar}
The covariance structure of the diagnostic will be based on the \textit{squared exponential kernel}:
\begin{align}
\kappa\left(\mathbf{x}, \mathbf{z}\right) = \alpha^{-d}\exp\left[-\frac{\lVert \mathbf{x} - \mathbf{z}\rVert^2}{2\lambda^2}\right], \label{eq:sq_exp}
\end{align}
a common choice in BQ \citep[e.g][]{OHagan1991, Karvonen2018, Briol2019}. The hyperparameter $\alpha$ controls the \textit{precision} of the GP, serving as a scaling factor for its variance and for that of its integral. It is more common in literature to parameterize the kernel in terms of scale as opposed to precision, replacing $\alpha^{-d}$ in equation~\eqref{eq:sq_exp} with $\alpha^2$ \citep[e.g.][]{OHagan1991, Gunter2014}, but the practical difference between these choices is purely notational. The parameterization in equation~\eqref{eq:sq_exp} is the same as that used by \citet{Chkrebtii2016}, and the fact that $\alpha$ is raised to the power of $-d$ in equation~\eqref{eq:sq_exp} reflects their notion that the $d$-dimensional kernel can be viewed as a pointwise product of $d$ univariate kernels. The hyperparameter $\lambda$ is the \textit{length-scale}, which controls the size of fluctuations in GP values between distinct points \citep{Rasmussen2006}. In informal terms\footnotemark, $\lambda$ therefore controls the ``smoothness'' of the GP.
\footnotetext{In formal terms, a GP with squared exponential covariance kernel is infinitely differentiable, in the mean square sense, regardless of the smoothing parameter value \citep{Rasmussen2006}. ``Smoothness'' as informally used above simply means an absence of ``wiggles'' at small scales in functions sampled from the GP.}

The actual covariance function used in the diagnostic is a modification of equation~\eqref{eq:sq_exp} based on the function of interest $f$. It is
\begin{align}
C_0^x(\mathbf{x},\mathbf{z}) = f\left(\hat{\mathbf{x}}\right)^2 \det\left(-\mathbf{H}^{-1}\right)\kappa\left(\mathbf{T}^{-1}\mathbf{x}, \mathbf{T}^{-1}\mathbf{z}\right), \label{eq:sq_exp_mod}
\end{align}
where the transformation matrix $T$ was defined in Section \ref{sec:points}. Because $\lVert \mathbf{T}^{-1}\mathbf{x} - \mathbf{T}^{-1}\mathbf{z}\rVert^2 = (\mathbf{x}-\mathbf{z})^T(-\mathbf{H})(\mathbf{x}-\mathbf{z})$, the prior covariance of the GP at distinct points depends on the distance between these points in a linear transformation of Euclidean space, with the transformation depending on the ``curvature'' of $\log f$ at $\hat{\mathbf{x}}$. Equivalently, the prior GP covariance function in equation~\eqref{eq:sq_exp_mod} is a (scaled) \textit{Mahalanobis kernel} \citep{Abe2005}.

\subsection{Choice of measure} \label{sec:measure}
In Section \ref{sec:pn}, we used an importance re-weighting trick to express $F$ as an integral w.r.t.\ a Gaussian measure $G$. \citet{OHagan1991} and \citet{Kennedy1998} considered BQ for $r = f/g$ with a constant GP prior mean and noted that results would be most accurate if the density $g$ closely approximated the shape of $f$, i.e.\ if $r$ was roughly constant. The latter noted an analogy with importance sampling (IS), in which $F$ is also modelled as the integral of $r$ w.r.t.\ G and the shape of $g$ should match that of the integrand \citep[e.g.][]{Yuan2007}. Although our GP prior mean (equation~\eqref{eq:prior_mean}) is not constant, we still found in preliminary experiments that $g$ had to be a fairly good ``fit'' to $f$ in order for the diagnostic to behave reasonably. Within the convenient class of Gaussian measures, remarks by O'Hagan and Kennedy suggest that $g$ proportional to the Gaussian approximation to $f$, i.e.\ $G = \mathcal{N}\left(\hat{\mathbf{x}}, -\mathbf{H}^{-1}\right)$, would be a reasonable ``starting point''. The measure ultimately used for the diagnostic is a slight modification of this:
\begin{align}
G = \mathcal{N}\left(\hat{\mathbf{x}}, -\gamma^2 \mathbf{H}^{-1}\right), \label{eq:measure}
\end{align}
where the new hyperparameter $\gamma > 0$ controls the ``spread'' of $G$ and will be discussed in Section \ref{sec:cal}.

% It turns out that the specification of $G$ is quite important in practice: preliminary experiments showed that a ``default'' choice of a standard multivariate normal distribution \citep[as in][]{Karvonen2018} could give disastrously inaccurate results depending on $f$.
%
%
% SOMETHING LIKE THIS:
% Despite not being stationary like O'Hagan and Kennedy, we found in preliminary experiments that the choice of measure did have an effect on results, and that the quality of these results improved when using their suggestion to take the measure proportional to the Gaussian approximation of f
% Note that no other combination of kernel, meausre, and interrogation scheme I've looked at gives the same invariance
%
%Finally, recall that a re-weighting trick equation~\eqref{eq:int_weight} is used so that the integral $F$ is reframed as the integral of $g = f/\mathrm{d}G$ w.r.t.\ the Gaussian measure $G$. We take $G = \mathcal{N}\left(\hat{\mathbf{x}}, -\gamma^2 \mathbf{H}^{-1}\right)$, where $\gamma > 0$ is another hyperparameter controlling the ``spread'' of this measure. Note that when $\gamma = 1$, the density $\mathrm{d}G$ is proportional to the Gaussian approximation to $f$. Intuitively (and practically, as we will explain later), it makes sense for the underlying measure to align with the ``shape'' of $f$ in this way.

\subsection{Invariance of diagnostic behaviour} \label{sec:invariance}
At first glance, it may seem that these function-specific design choices are antithetical to the intended ``one-size-fits-all'' nature of the diagnostic. On the contrary, our design ensures a few kinds of advantageous ``invariance''. Recall that the interrogation points are obtained from the function-agnostic preliminary grid as $\mathbf{s}_i = \mathbf{T}\mathbf{s}^*_i + \hat{\mathbf{x}}$, $i \in \{ 1, \dots, n\}$. Plugging any two interrogation points $\mathbf{s}_i, \mathbf{s}_j$ into equation~\eqref{eq:sq_exp_mod} therefore gives $C^x_0(\mathbf{s}_i, \mathbf{s}_j) \propto \kappa(\mathbf{s}^*_i, \mathbf{s}^*_j)$. Note also that analogous results can be shown to hold for the integral terms\footnotemark{} in Equations~\eqref{eq:int_mean} -- \ref{eq:int_variance} and for the prior mean interrogations $\mathbf{m}_0^x\left(\mathbf{S}\right)$. %the posterior mean and variance do \textit{not} depend on the location of $\hat{\mathbf{x}}$, and only depend on $G$ through scaling factors. Thus,
Therefore, in principle the interrogations should provide the same quality and quantity of ``information'' for \textit{any} $f$. Now, recall that the diagnostic rejects the LA for $f$ iff it is not contained in the central 95\% interval of the integral posterior, i.e.\ iff $L(f) \notin \left(m_1 - 1.96\sqrt{C_1}, m_1 + 1.96\sqrt{C_1}\right)$. Note that $\sqrt{C_1}$ is equal to $L(f) \propto f\left(\hat{\mathbf{x}}\right)\sqrt{\det\left(-\mathbf{H}^{-1}\right)}$ times a factor depending only on $\mathbf{S}^*$ and the hyperparameters $(\lambda, \alpha, \gamma)$ (by equations~\eqref{eq:int_variance} and~\eqref{eq:sq_exp_mod}); similarly, $m_1$ is equal to $L(f)$ times a factor depending only on $\mathbf{S}^*$, the hyperparameters, and the ``normalized'' function values $\mathbf{f}\left(\mathbf{S}\right)/f(\hat{\mathbf{x}})$ (by equations~\eqref{eq:prior_mean},~\eqref{eq:int_mean}, and the definition of $r$). Thus, the necessary and sufficient condition for rejection does not depend on the actual values of $\hat{\mathbf{x}}$, $f\left(\hat{\mathbf{x}}\right)$, and $\det\left(-\mathbf{H}^{-1}\right)$: \textit{it is invariant to any scaling of the function or affine transformation of its domain}. More formally, for a fixed set of hyperparameters, the diagnostic rejects the LA when applied to $f$ iff it rejects the LA when applied to any function of the form $f_\mathrm{Trans}: \mathbf{x} \mapsto af\left(\mathbf{Ax} + \mathbf{b}\right)$ with $a > 0$, $\mathbf{A} \in \mathbb{R}^{d \times d}$ with $\det{\mathbf{A}} \neq 0$, and $\mathbf{b} \in \mathbb{R}^d$. The only way in which $f$ affects the result of the diagnostic is through the \textit{relative} differences between its values at the interrogation points and those of its Gaussian approximation. Because the diagnostic seeks only to determine whether $f$ is ``sufficiently Gaussian in shape'', this is precisely the appropriate behaviour for it to have.
\footnotetext{To see this,
%note that the density $g$ has a multiplicative factor of $\sqrt{\det{\left(-H\right)}} = \left\lvert\det{\left(T^{-1}\right)}\right\rvert$, and
integrate equation~\eqref{eq:sq_exp_mod} by substitution. This is another reason why the choice of measure in equation~\eqref{eq:measure} makes sense.}

Note the ``standardized'' design developed in Sections \ref{sec:points}--\ref{sec:measure} is not without precedent in the BQ literature. For instance, \citet{Sarkka2015} adopted the idea of \textit{stochastic decoupling} from sigma-point methodology: to integrate a function $r$ against some Gaussian measure $\mathcal{N}(\mathbf{\mu}, \mathbf{P})$, they placed a GP prior with the standard squared exponential covariance kernel (equation~\eqref{eq:sq_exp}) on the function $r_\mathrm{Trans}: x \mapsto r(\mathbf{\mu} + \sqrt{\mathbf{P}}\mathbf{x})$ and used a standardized set of ``unit'' interrogation points. Such an approach is essentially equivalent (possibly up to variance scaling factors) to our design; indeed, the authors made note of its invariance to affine transformations. However, their main interest was in deriving BQ-based methods for filtering and smoothing in nonlinear SSM's, in which $\mathbf{\mu}$ and $\mathbf{P}$ are computed for each necessary integral according to their algorithms \citep{Sarkka2015}.
%To our knowledge, the use of the function-dependent factors $\hat{\mathbf{x}}$ and $H$ (and the assumption that they are known \textit{a priori}) to ``standardize'' the design is unique to the present work.

\section{Hyperparameter calibration} \label{sec:cal}
It remains to select values for $(\lambda, \alpha, \gamma)$. As discussed above, the design of the interrogation grid and covariance kernel serve to ``standardize'' the input and output scales of the GP, so it is not necessary to consider these factors when setting the hyperparameters. Indeed, for a given dimension $d$ and preliminary grid $\mathbf{S}^*$, the same hyperparameter values should be used for \textit{any} $f$ to ensure the aforementioned diagnostic invariance. Recall from the beginning of Section \ref{sec:design} that the intent is to test ``good-enough-ness of fit'': the diagnostic should reject the LA for functions with a substantially non-Gaussian shape, but should \textit{not} be so ``powerful'' that it rejects functions which are close enough to Gaussian. With this in mind, we propose to set the hyperparameters in a somewhat heuristic way based on a predetermined \textit{calibration} or \textit{test function} $\tau$. Such a function should have a shape fairly close to Gaussian in order to serve as the ``edge case'' for the diagnostic. Specifically, given a preliminary grid $\mathbf{S}^*$ and test function $\tau$, the hyperparameters for the $d$-dimensional diagnostic should be set such that the following conditions are met when the diagnostic is applied to $\tau$.
\begin{enumerate}
\item[(1)] The LA $L(\tau)$ should be on the boundary of the rejection region (i.e.\ equal to one of the endpoints of the 95\% central interval for the integral posterior); and
\item[(2a)] the discrepancy between $\tau$ and the ``un-weighted'' posterior GP mean, $m_1^x\cdot g$, should be as small as possible throughout the domain; or at the very least
\item[(2b)] the posterior integral mean $m_1$ should be as close as possible to the true integral of $\tau$.
\end{enumerate}
Either version of the second condition should ensure that the diagnostic is reasonably accurate when applied to $\tau$. Of course, accurate estimation is still an ancillary goal in general, but at the very least it should be achieved for the test function to ensure that the diagnostic uses interrogations in a sensible way. Condition (2a) is the more desirable version since it directly targets the shape of the function and also implies (2b) by design, but in high dimensions with large interrogation grids it may only be possible to ensure that (2b) is met (see Section \ref{sec:highdim}). The first condition establishes $\tau$ as the ``borderline'' function: any function that is ``less Gaussian'' will have its LA rejected, and any function ``at least as Gaussian'' will not. To see this, consider the normalized posterior ``correction term''\footnotemark\footnotetext{To avoid any possible confusion, it should be reiterated that all of the quantities in these definitions - technically depend on the integrand through the constructions detailed in Sections \ref{sec:pn}--\ref{sec:measure}. More accurate notation would reflect this explicitly, but such notation would be cumbersome.}
\begin{align}
\Delta(f) :=  \frac{\sqrt{\det\left(-\mathbf{H}\right)}}{f\left(\hat{\mathbf{x}}\right)}\left[\int_{\mathbb{R}^d}\mathbf{C}_0^x(\mathbf{x},\mathbf{S})\mathrm{d}G\left(\mathbf{x}\right)\right]^T\left[\mathbf{C}^x_0(\mathbf{S}, \mathbf{S})\right]^{-1}\left(\mathbf{r}(\mathbf{S}) - \mathbf{m}_0^x(\mathbf{S})\right), \label{eq:correction}
\end{align}
which, as per equation~\eqref{eq:int_mean}, is (up to the scaling factors in front) the difference between the prior and posterior integral means when the diagnostic is applied to a function $f$. It can be shown that the rejection criterion for the diagnostic is equivalent to $f\left(\hat{\mathbf{x}}\right)\sqrt{\det\left(-\mathbf{H}^{-1}\right)}\left\lvert \Delta(f) \right\rvert > 1.96\sqrt{C_1}$. Recall from Section \ref{sec:invariance} that $C_1$ only depends on $f$ through scaling factors $f\left(\hat{\mathbf{x}}\right)^2$ and $\det\left(-\mathbf{H}^{-1}\right)$, so the rejection criteria is equivalent to $\left\lvert\Delta(f)\right\rvert > \epsilon$, where the number $\epsilon > 0$ depends only on $\mathbf{S}^*, \lambda, \alpha$, and $\gamma$. Now, to meet condition (1) for the test function $\tau$ is to have $\left\lvert \Delta(\tau) \right\rvert = \epsilon$. Therefore, with this calibration scheme a function $f$ will have its LA rejected iff $\left\lvert \Delta(f) \right\rvert > \left\lvert \Delta(\tau) \right\rvert$. Again, all that matters are the \textit{relative differences} between a function and its Gaussian approximation at the interrogation points --- specifically, whether the weighted sum of these as given by equation~\eqref{eq:correction} (with the weights depending on $\mathbf{S}^*, \lambda$, and $\gamma$) is larger in magnitude than it is for the predetermined ``borderline Gaussian'' $\tau$.

A natural choice for a test function is the density of a $d$-dimensional multivariate Student's $t$ distribution with $\nu$ degrees of freedom, mean at the origin, and scale matrix equal to the identity. Denote this density by $\tau_{\nu, d}$, so
\begin{align}
\tau_{\nu,d}(\mathbf{x}) = \frac{\Gamma\left(\frac{\nu+d}{2}\right)}{\Gamma\left(\frac{\nu}{2}\right)\sqrt{\nu\pi}}\left(1+\frac{\lVert \mathbf{x} \rVert^2}{\nu}\right)^{-\frac{\nu+d}{2}}, \label{eq:mult_t}
\end{align}
and note that it has heavier tails than a $d$-dimensional Gaussian density, so the LA, given by the formula
\begin{align}
L\left(\tau_{\nu, d}\right) = \left(\frac{2}{\nu + d}\right)^{\frac{d}{2}}\frac{\Gamma\left(\frac{\nu+d}{2}\right)}{\Gamma\left(\frac{\nu}{2}\right)}, \label{eq:mult_t_lap}
\end{align}
underestimates the true integral (which is always equal to 1). However, $\tau_{\nu, d}$ approaches a standard multivariate normal density in the limit $\nu \to \infty$, and therefore $L\left(\tau_{\nu, d}\right) \to 1$ as well. Therefore, for some large value of $\nu$, the shape of $\tau_{\nu, d}$ may be said to be ``sufficiently Gaussian'' to warrant non-rejection of the LA. Denote such a value by $\nu_d$ to reflect the fact (discussed further in Section \ref{sec:highdim}) that the specific choice of test function should depend on the dimension $d$. One option that works reasonably well
%(at least, in low and moderate dimensions)
is to let $\nu_d$ be the smallest integer such that $L\left(\tau_{\nu_d, d}\right) \geq 0.95$. The densities of multivariate $t$ variables with more than $\nu_d$ degrees of freedom are close enough in shape to Gaussians that their Laplace approximations are within 5\% of the true integral value; conversely, those with lower degrees of freedom have heavier tails and LA's that underestimate the true integral by over 5\%.

With the family of test functions established, it is now possible to discuss how one may set the hyperparameters to satisfy the conditions listed above. First note that the precision parameter $\alpha$ does not actually affect the posterior mean; as a scaling factor, it serves only to ensure that condition (1) is met. Thus, it suffices to find good values for $\lambda$ and $\gamma$, after which $\alpha$ can simply be chosen to scale the posterior variance $C_1$ such that $\left\lvert \Delta(\tau_{\nu_d,d}) \right\rvert = \epsilon$.

The fact that $\lambda$ affects the shape of the GP mean is obvious since, as noted in Section \ref{sec:covar}, it determines the ``smoothness'' of functions sampled from the GP and is therefore a ``shape parameter'' in some sense. What is perhaps more surprising is the effect of $\gamma$, the scaling factor for the underlying measure $G$. Recall from Section \ref{sec:measure} that $G$ is analogous to the proposal distribution in IS. It is well-known that the performance of an importance sampler will be poor if the density $g$ has lighter tails than $f$, and it is therefore better to err on the side of caution by taking $g$ to have slightly heavier tails \citep[e.g.][]{Yuan2007}. In our context, this corresponds to setting $\gamma$ slightly larger than 1, and in our experiments we use a value of
\begin{align}
\gamma = &\sqrt{1.5\frac{\nu_d + d}{\nu_d + d - 3}}. \label{eq:gamma}
\end{align}
% \mathrm{Var}\left[Y_1 \mid Y_2 = 0, \dots, Y_d = 0 \right] &= \mathrm{Var}\left[X_1 \mid  X_2 = 0, \dots, X_d = 0 \right] \nonumber
%\end{align}
The heuristic motivation for this choice is as follows. Consider $d$-dimensional random vectors $\mathbf{Y} \sim \tau_{\nu_d, d}$ and $\mathbf{X} \sim g$, where $g = g(\tau_{\nu_d,d})$ is the density corresponding to equation~\eqref{eq:measure} for the choice of function $f = \tau_{\nu_d, d}$. The $\gamma$-value given by equation~\eqref{eq:gamma} ensures that $1.5\times\mathrm{Var}[Y_1 \mid Y_2 = 0, \dots, Y_d = 0 ] = \mathrm{Var}[X_1 \mid  X_2 = 0, \dots, X_d = 0 ]$ --- in words, the univariate conditional densities (with all other coordinates fixed at the origin) of the $t$ distribution used for calibration have variance equal to two thirds of those of the ``approximating Gaussian density'' $g$ \citep{Kennedy1998}. Here the analogy with IS becomes somewhat strained, as it can be shown that \textit{any} Gaussian proposal distribution will result in an importance sampler with infinite variance when applied to a $t$ density. In fact, taking $G$ itself as a $t$ distribution is often a good choice in IS due to the heaviness of the tails \citep[][and references therein]{Tokdar2009}. \citet{Pruher2017} considered this choice of $G$ in BQ, but noted that the kernel integrals in equations~\eqref{eq:int_mean}--\eqref{eq:int_variance} would not have closed forms. For computational convenience we will retain our choice of a Gaussian measure, but note that, unlike IS, the posterior variance of the integral is still guaranteed to be finite here.

\subsection{Calibrating in two dimensions} \label{sec:2d}
Using these ideas, we will now demonstrate how calibration can work for the diagnostic in $d = 2$ dimensions. As with many of the Figures in this manuscript, the results were obtained using MATLAB, version 2024a \citep{MATLAB}. The test function will be a bivariate $t$ density with $\nu_2 = 38$ degrees of freedom, as $L\left(\tau_{38, 2}\right) = 0.95$. The preliminary interrogation grid $\mathbf{S}^*$ will consist of evenly-spaced points in a ``cross-shaped'' formation ``on the axes'' of $\mathbb{R}^2$:
\begin{align}
\mathbf{S}^* = \left\{(0, 0)\right\} \cup \left\{\pm m\mathbf{e}_i: m \in \{1, 2, 3\},\ i \in \{1, 2\}\right\}, \label{eq:2d_grid}
\end{align}
where $e_i$ is the $i^\mathrm{th}$ standard basis vector of $\mathbb{R}^2$. Such ``cross-shaped grids'' are appealing, at least in low dimensions, because the number of points $n$ scales linearly with $d$. Here, we have $n = 13$.

In order to heuristically understand how hyperparameter choices affect the behaviour of the diagnostic, it will be useful to plot the difference between the test function $\tau_{38,2}$ and the ``un-weighted'' GP posterior mean $m^x_1\cdot g$ for various $(\lambda, \gamma)$-values. Note that the ``optimal'' hyperparameters will depend on the dimensionality of the domain, the specific test function used, and the preliminary grid chosen. In particular, if one wishes to use the diagnostic in 2 dimensions with a different preliminary grid from the one considered here, it should not necessarily be assumed that the $\lambda$ value given below is suitable for the new grid.

Choosing $\gamma$ according to equation~\eqref{eq:gamma} with $d = 2$ and $\nu_d = 38$ results in a value of $\gamma = 1.2734$. In this low-dimensional setting with a small interrogation grid, it is possible to crudely approximate an analytic method to find an ``optimal'' $\lambda$: given the aforementioned $\gamma$-value, we approximate the ``$L^2$ error'' $\int_{\mathbb{R}^2} \left(m^x_1(\mathbf{x})g(\mathbf{x}) - \tau_{38, 2}(\mathbf{x})\right)^2\mathrm{d}\mathbf{x}$ and its derivative w.r.t.\ $\lambda$ by simple Riemann sums over the grid of points $\left\{-10, -9.99, -9.98, \dots, 9.99, 10\right\}^2$. This approximate error is then minimized w.r.t.\ $\lambda$ using the BFGS algorithm as implemented in the \texttt{fminunc} function in the \texttt{MATLAB Optimization Toolbox} \citep{MatlabOTB}, resulting in a value of $\lambda = 4.2241$. %Note: there are 4 "standard" references for the BFGS algorithm. Should I include these, or will the Optimization Toolbox citation suffice?

\begin{figure}
\centering{$\lambda = 4.2241$, $\gamma = 1.2734$, $\alpha = 0.023142$}
\centering{\includegraphics[width=\textwidth, trim = {6cm 0 1cm 0}, clip]{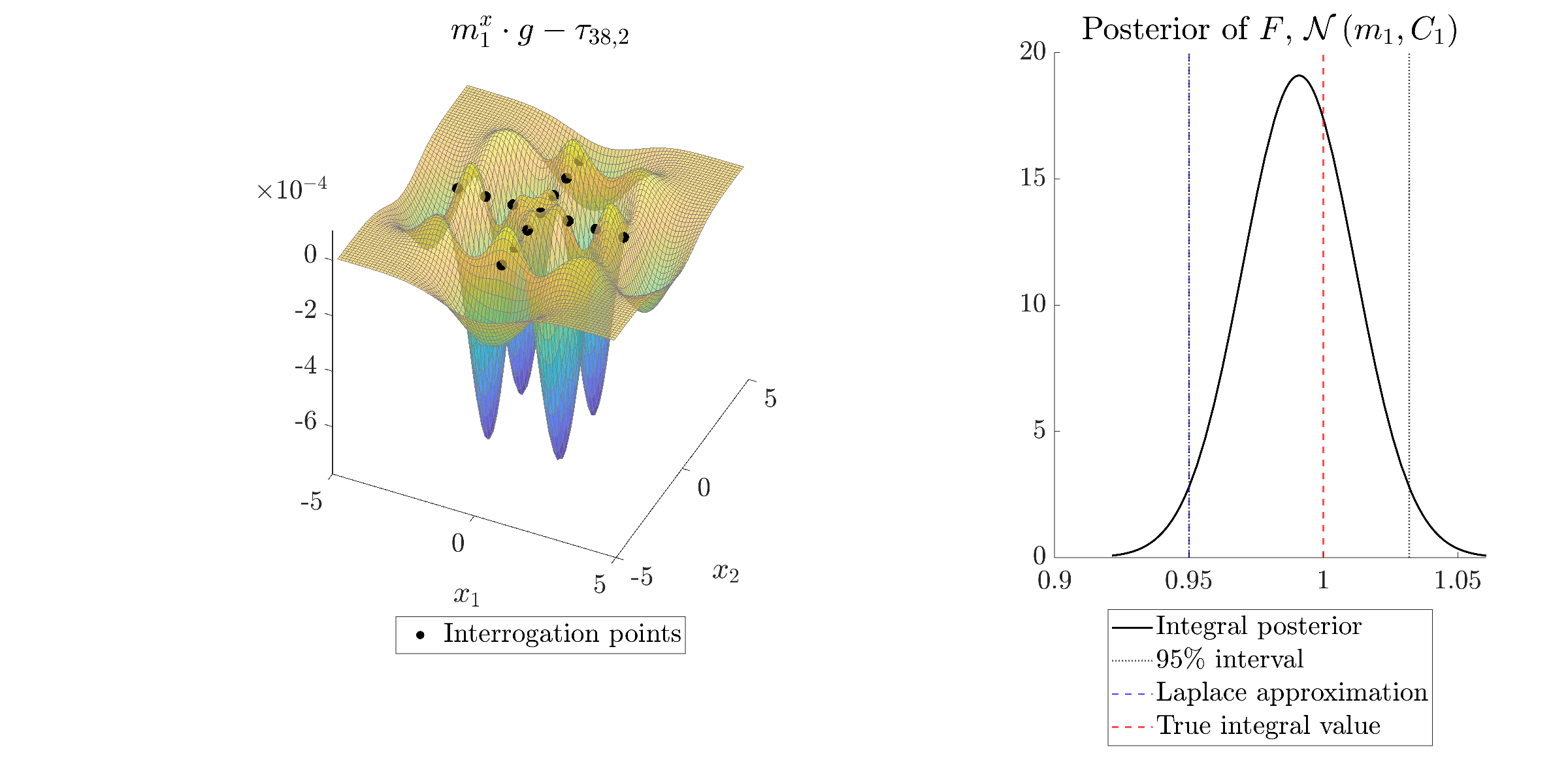}}
\caption{Results for the diagnostic applied to the 2-dimensional test function $\tau_{38,2}$, with an ``optimal'' $\lambda$, $\gamma$ obtained from equation~\eqref{eq:gamma}, and $\alpha$ set to ensure that the LA is on the boundary of the ``rejection region''. Left: the difference between the un-weighted posterior GP mean and the true function. Right: the posterior distribution for the integral $F$.} \label{fig:2d_opt}
\end{figure}

\begin{figure}
\centering{$\lambda = 0.0729$, $\gamma = 1.2734$, $\alpha = 25.2372$}
\centering\includegraphics[width=\textwidth, trim = {5.2cm 0 1cm 0}, clip]{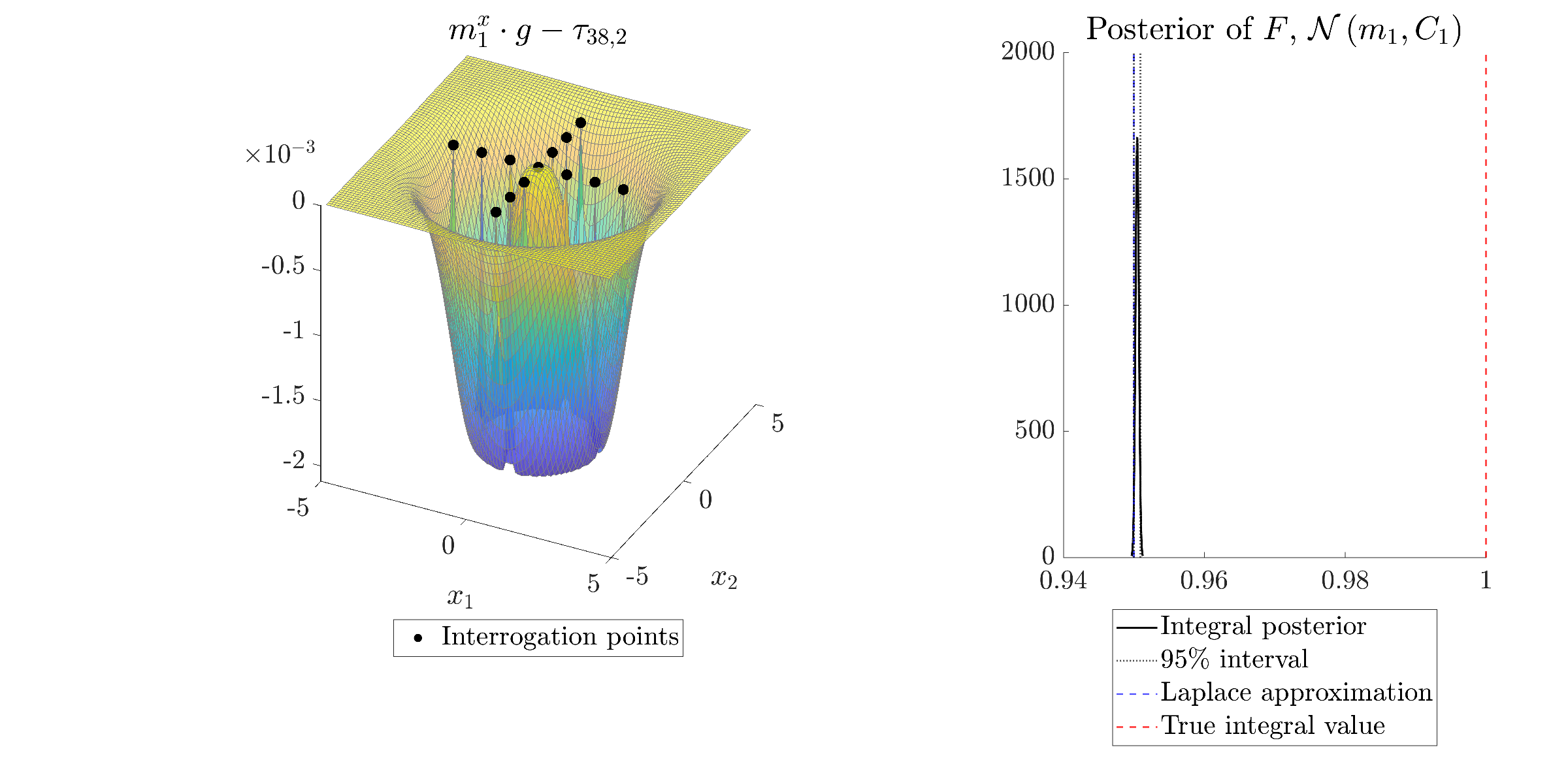}
\caption{Results for the diagnostic applied to $\tau_{38,2}$ with a low $\lambda$-value, $\gamma$ obtained from equation~\eqref{eq:gamma}, and $\alpha$ set to ensure that the LA is on the boundary of the ``rejection region''. Note the spikes created by ``undersmoothing''.}\label{fig:2d_lambda_low}
\end{figure}

\autoref{fig:2d_opt} shows results for the diagnostic applied to $\tau_{38,2}$ with these design choices. The difference $\left(m^x_1 \cdot g - \tau_{38,2}\right)$ is very small among the lines defined by the interrogation grid, but there are deep valleys centered around the ``main diagonals'' of the plane and within the boundaries of the interrogation grid. Since the heavy-tailed $t$ density is larger than its Gaussian approximation in these regions, it is clear that there is not much difference between the prior and posterior GP means there. The interrogation points are too far from these regions to exert much influence on the posterior mean there - in this respect, one may say that the GP is failing to \textit{interpolate} to these areas. A more mathematical explanation of this behaviour can be extracted from equation~\eqref{eq:mean}, the definition of $m_1^x$. By this definition, it holds that $m_1^x(\mathbf{s})g(\mathbf{s}) = fAA(\mathbf{s})$ for any function $f$, interrogation point $\mathbf{s}$, and combination of hyperparameter values. However, at any other (non-interrogation) point $\mathbf{x}$, the extent to which $m_1^x(\mathbf{x})$ updates from the prior GP mean $m_0^x(\mathbf{x})$ is determined by the ``weights'' $\mathbf{C}_0^x(\mathbf{x}, \mathbf{S})^T[\mathbf{C}_0^x(\mathbf{S}, \mathbf{S})]^{-1}$. These weights tend to decrease in magnitude as $x$ moves away from the points in $\mathbf{S}$, to an extent determined by $\lambda$ and $\gamma$. When $\lambda$ is small, there is almost no prior dependence between GP values at distinct points, so these weights are close to zero for $\mathbf{x} \notin \mathbf{S}$. This can be seen in \autoref{fig:2d_lambda_low}: the posterior GP mean is forced to equal $\tau_{38,2}$ at the interrogation points, but everywhere else it is virtually unchanged from the prior mean $m_0^x$). Thus, in this case $m_1$ is very close to the prior value $m_0 = L\left(\tau_{38,2}\right) = 0.95$. In contrast, the ``optimal'' $\lambda$-value results in a posterior integral estimate of $m_1 = 0.99095$, quite close to the true value of 1. Note that in each case, the integral of $\left(m^x_1 \cdot g - \tau_{38,2}\right)$ (the surface in the left plot) over $\mathbb{R}^2$ is equal to the difference between $m_1$ and the true integral (in the right plot, the horizontal distance between the peak of the bell curve and the red line). As mentioned above, $\alpha$ is chosen to ensure that the test function is on the boundary between rejection and non-rejection, resulting in a posterior variance of $C_1 = 4.3653\times10^{-4}$ for the ``optimal'' $\lambda$ and $5.7369\times10^{-8}$ for the lower one.


The effect of $\gamma$ is less easily explained than that of $\lambda$. In fact, their effects counterbalance each other to some degree: we found that it was still possible to approximate an ``optimal'' $\lambda$ with the method described above even for different fixed values of $\gamma$, with lower $\gamma$-values resulting in higher required $\lambda$-values and vice-versa. In principle, this suggests that the diagnostic will not be too sensitive to the use of different $\gamma$-values, since any possible negative effect on its performance could be mitigated by adjusting $\lambda$ in the opposite direction. However, there is a limit to this in practice, and $\gamma$-values that are either too low or too high can still be problematic. With a lower value of $\gamma = 1$, it became difficult to find an optimal $\lambda$, as the BFGS algorithm was quite sensitive to the choice of initial value. Although the results of differently-initialized BFGS runs were not consistent with each other, they all resulted in final $\lambda$-values over 9. At length-scales this large, the \textit{Gram matrix} $\mathbf{C}^x_0(\mathbf{S}, \mathbf{S})$ is poorly conditioned (for instance, with $\mathbf{S}^*$ given by expression~\eqref{eq:2d_grid}, its reciprocal condition number is $7.7885\times10^{-14}$ when $\lambda = 9$, as opposed to $7.1579\times10^{-10}$ when $\lambda = 4.2241$), so numerical stability becomes a concern. Furthermore, even with $\lambda$-values this high, the posterior integral mean $m_1$ was around 0.986: not as close to 1 as it was with the slightly larger $\gamma$-value and its ``optimal'' $\lambda$. The fact that these difficulties exist for $\gamma = 1$ is noteworthy since this corresponds to using an integrating measure whose density is proportional to the Gaussian approximation to the true function.

\begin{figure}
\centering{$\lambda = 1.3$, $\gamma = 3$, $\alpha = 1.39$}
\centering\includegraphics[width=\textwidth, trim = {5.2cm 0 1cm 0}, clip]{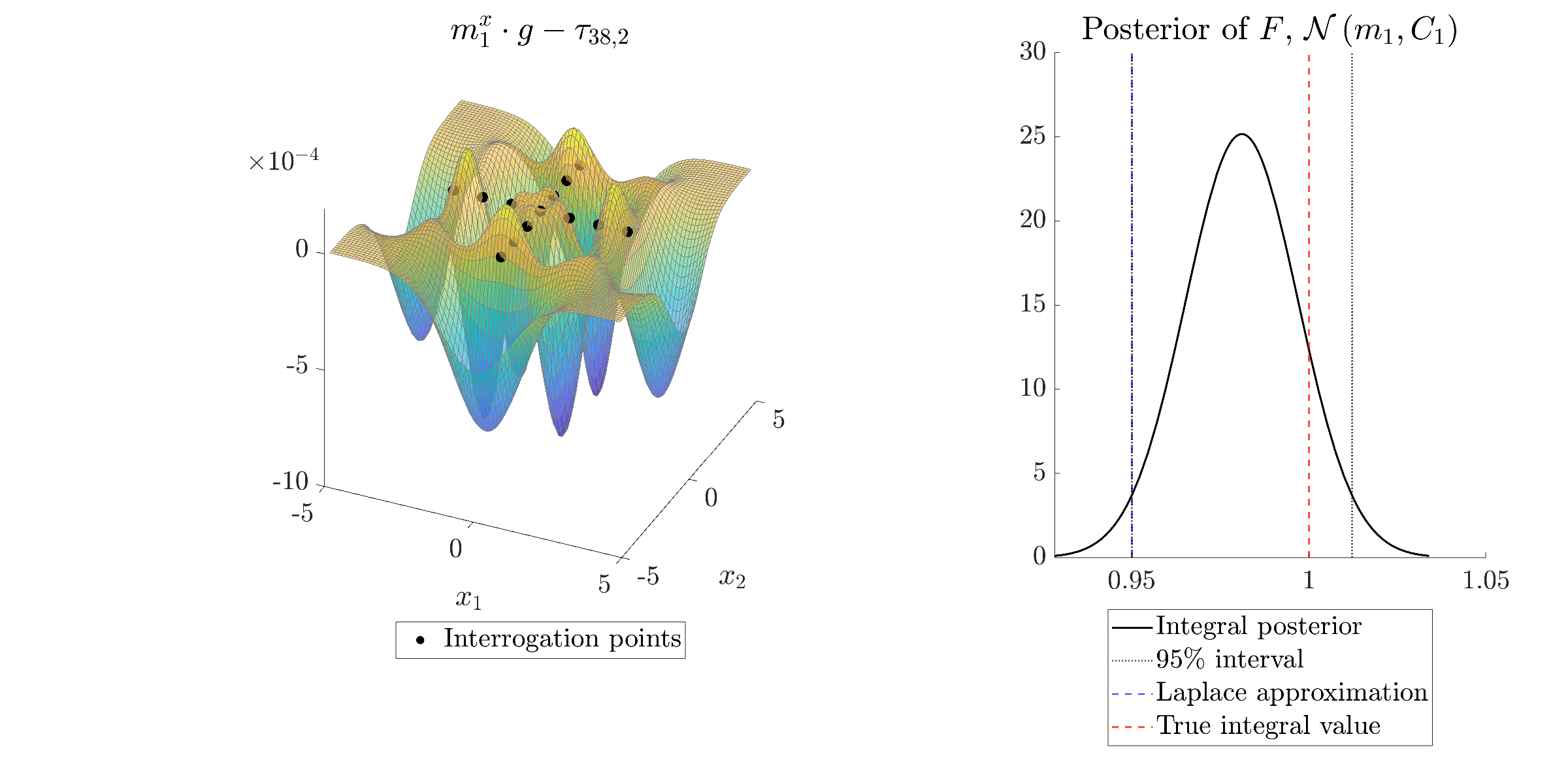}
\caption{Results for the diagnostic applied to $\tau_{38,2}$ with a high $\gamma$-value, $\alpha$ set to ensure that the LA is on the boundary of the ``rejection region'', and a $\lambda$-value that is only slightly larger than the approximate ``$L^2$ optimum'' for this $\gamma$-value (which, in this case, was $\lambda = 1.1953$). Contrast with \autoref{fig:2d_opt} to see the excessive sensitivity to $\lambda$ caused by a high $\gamma$-value.}\label{fig:2d_oversmooth}
\end{figure}

Concerns about numerical accuracy do not exist with an even larger $\gamma$-value, as the corresponding optimal $\lambda$-value will be smaller and the Gram matrix will therefore be better conditioned. However, sensitivity becomes a problem in this situation: when $\gamma$ is high, even a relatively small deviation from the optimal $\lambda$ can change the diagnostic's behaviour quite dramatically. This will be of particular concern in higher dimensions, in which it is not viable to approximate and optimize the $L^2$ error numerically. In the current 2-dimensional setting, with $\gamma = 3$, the approximately-optimal $\lambda$-value is 1.1953, and the results with these hyperparameters (not shown) are fairly similar to those in \autoref{fig:2d_opt}. A modest increase to $\lambda = 1.3$ creates a noticeably different outcome, as shown in \autoref{fig:2d_oversmooth}. The ``interpolation valleys'' seen in \autoref{fig:2d_opt} are slightly smaller in size, as the larger length-scale increases dependence between distinct points in the GP, thereby allowing the interrogations to exert more influence at faraway points. However, this slight improvement in interpolation comes at a cost: undesirable \textit{extrapolation} effects due to oversmoothing. Indeed, in all four directions just beyond the extremal interrogation points, $m_1^x$ dips well below the true function $\tau_{38,2}$. As a result, $m_1 = 0.98108$ is farther from the true integral than it was with the hyperparameter values in \autoref{fig:2d_opt}. Oversmoothing causes the weights $\mathbf{C}_0^x(\mathbf{x}, \mathbf{S})^T[\mathbf{C}_0^x(\mathbf{S}, \mathbf{S})]^{-1}$ to have unpredictable effects at $x$ beyond the boundaries of the interrogation grid, depending on the spread and density of $\mathbf{S}$ as well as the shape of the integrand. In some cases, the ``extrapolation valleys'' seen in \autoref{fig:2d_oversmooth} may be replaced by large ``hills'', causing $m_1$ to significantly overestimate the value of $F$ (not shown). It is now clear that the original hyperparameter values in \autoref{fig:2d_opt} provide the best ``tradeoff'', balancing the interpolation errors of undersmoothing with the extrapolation errors of oversmoothing.

\section{Example: a banana-shaped function} \label{sec:banana}
\begin{figure}
\centering\includegraphics[width=\textwidth, trim = {4.5cm 0.5cm 0.5cm 1.5cm}, clip]{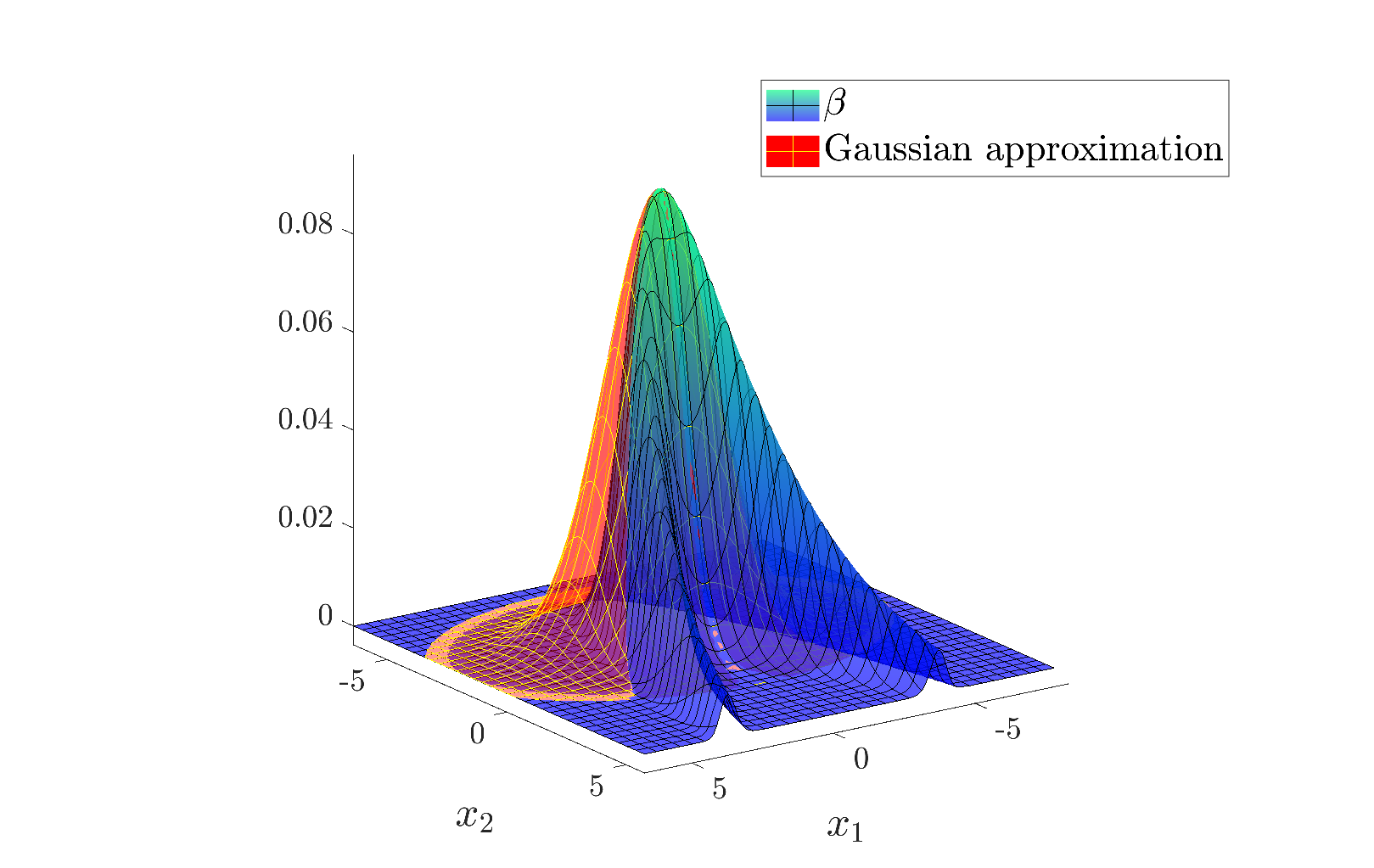}
\caption{A two-dimensional ``banana-shaped'' function alongside its Gaussian approximation.}\label{fig:banana}
\end{figure}

\begin{figure}
\centering\includegraphics[width=\textwidth, trim = {5cm 0 1cm 0}, clip]{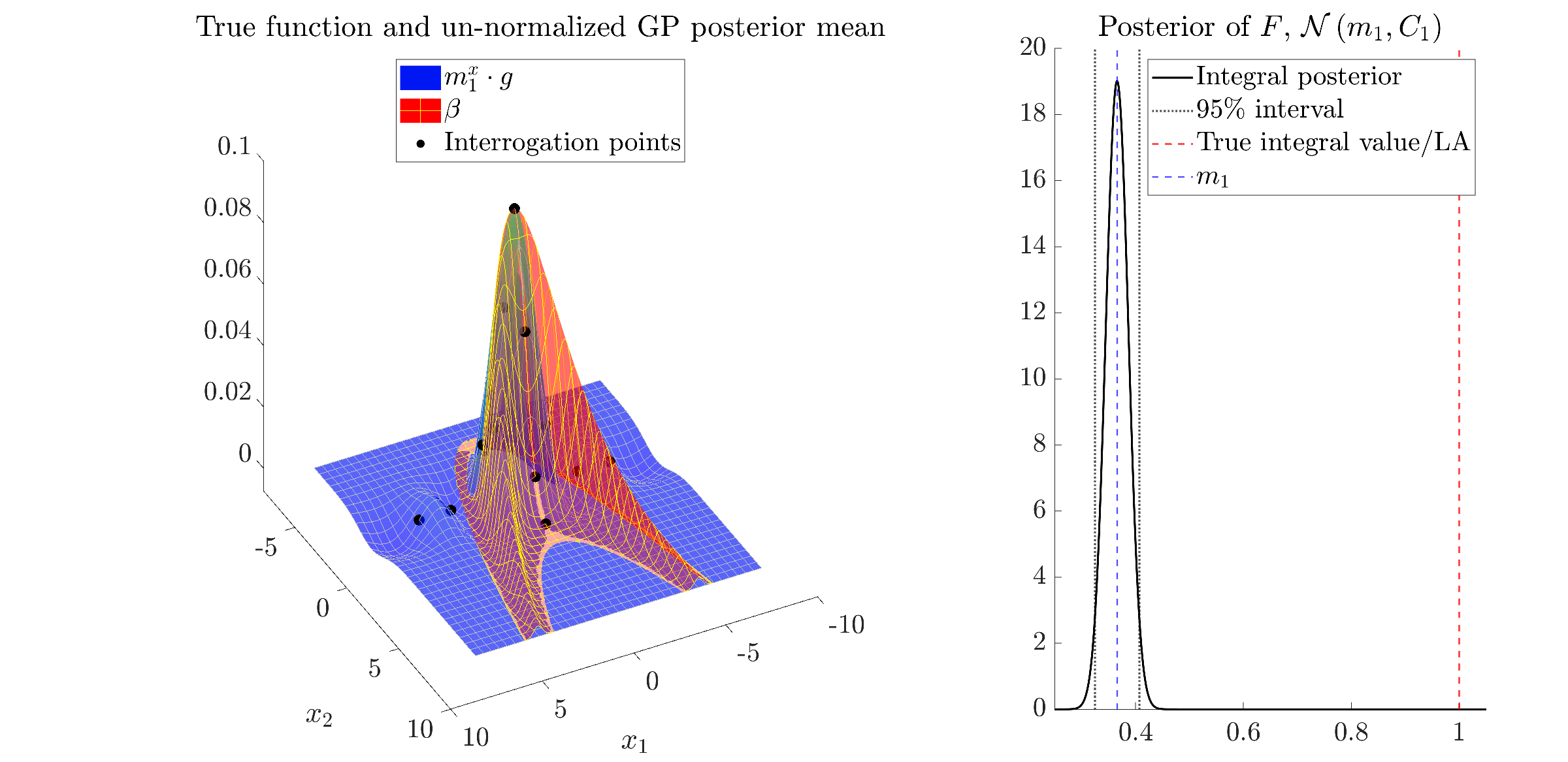}
\caption{Results from applying the diagnostic to the two-dimensional banana-shaped function, using the same design choices as in \autoref{fig:2d_opt}. Note that the colours in the left plot are reversed from those in \autoref{fig:banana} for easier visualization.}\label{fig:post_banana}
\end{figure}

In a paper on MCMC algorithms, \citet{Haario1999} considered  a function with ``banana-shaped'' %or ``boomerang-shaped''
contours, defined by ``twisting'' one coordinate of a Gaussian density. Letting $\varphi\left(\cdot; \Sigma\right)$ denote a bivariate Gaussian density with mean at the origin and covariance matrix $\Sigma$, the version of the function used here is
\begin{align*}
\beta(\mathbf{x}) := \varphi\left(x_1, x_2 - \frac{1}{2}\left(x_1^2 - 3\right);
\begin{pmatrix}
3 & 0 \\
0 & 1
\end{pmatrix}
\right).
\end{align*}
It turns out that the Laplace approximation is true for this function: $L\left(\beta\right) = \int_{\mathbb{R}^2} \beta(\mathbf{x}) \mathrm{d}\mathbf{x} = 1$. As discussed in Section \ref{sec:framework}, this may be viewed as ``coincidence'', as it is clear from \autoref{fig:banana} that $\beta$ is not well-approximated by a Gaussian shape. In this way, the function $\beta$ represents an interesting test case for the diagnostic: although its LA is technically valid, it is \textit{not} ``Gaussian enough'' and should therefore be rejected. Indeed, with the preliminary interrogation grid (expression~\eqref{eq:2d_grid}) and corresponding (approximately) ``optimal'' hyperparameters (see \autoref{fig:2d_opt}), this is precisely what the diagnostic does, as shown in \autoref{fig:post_banana}. The un-weighted GP posterior mean now accurately captures the light tails of $\beta$ along the line $x_2 = 0$, although it does not capture the large ridges defining the ``banana'' shape since there are no interrogation points along these ridges. As a result, the posterior integral estimate $m_1$ is 0.3658 --- well below the true value and the LA. Note also that there are small oscillations between the interrogation points along the $x_1$-axis, perhaps signifying a small amount of oversmoothing. Finally, observe that the posterior variance is small enough to result in a rejection of the LA, which is well above the 97.5\% quantile for the posterior distribution of $F$. These design choices would certainly be poor ones if accurate integral estimation was the main goal. In this framework, however, they are clearly suitable --- the shape information captured by the diagnostic suggests that $\beta$ is not Gaussian enough to justify using the LA outright. In this type of scenario, a practitioner could subsequently employ a different method to estimate the integral. Presumably, they would then discover that the LA was correct all along --- but \textit{not} because of the quality of the Taylor approximation~\eqref{eq:taylor} underpinning its use.


\section{The diagnostic in high dimensions: considerations and applications} \label{sec:high}
\subsection{Overview} \label{sec:highdim}
The low-dimensional LA diagnostic experiments Sections \ref{sec:2d} and \ref{sec:banana} were useful for exposition, but ultimately our main interest is in applying the diagnostic to higher-dimensional functions. Unsurprisingly, for large dimensions $d$ it is more challenging to ensure that the diagnostic behaves well. Recall from Section \ref{sec:2d} that we found an approximately ``optimal'' length-scale $\lambda$ by minimizing a type of $L^2$ error associated with the calibration function $\tau_{\nu_d, d}$. This required the numerical approximation of an integral over $\mathbb{R}^d$, which is not computationally feasible in high dimensions (if it was, there would be no need for the LA or for this very diagnostic). It is also not viable to seek a closed-form expression for the $L^2$ error: doing so would, in turn, require an analytic expression for the inverse of the Gram matrix $\mathbf{C}^x_0\left(\mathbf{S}, \mathbf{S}\right)$, which will be prohibitively complicated for all but the smallest of interrogation grids. With respect to the conditions for hyperparameter calibration listed in Section \ref{sec:cal}, condition (2a) can be assessed with a heuristic visual approach for moderate dimensions $d > 2$: viewing a 2-dimensional ``slice'' of the difference $m^x_1\cdot g - \tau_{\nu_d,d}$ with $x_3, \dots, x_d$ all set to 0 (exploiting the symmetry of the $t$ density and the fact that its mode is at the origin), one can adjust $\lambda$ so as to make this difference appear as uniformly small as possible, attempting to balance issues with interpolation and extrapolation. Unfortunately, even this approach ceases to be viable when $d$ is large, so that Condition (2b) is all that can be ensured. The reasons for this depend on the structure of the preliminary grid $\mathbf{S}^*$; in turn, this structure should be chosen to mitigate the challenges that arise in high dimensions. More details on some possible choices are given below. We found in preliminary experiments that grids of the form in expression~\eqref{eq:2d_grid} --- that is, those with multiple evenly-spaced points along each axis --- did not work very well when generalized to higher dimensions. Note that, although the points along any given axis are equally spaced in such grids, the distances between points on \textit{different} axes will be larger. We conjecture that this variation in interrogation point distances becomes problematic in high dimensions as more axes and points are added.

Fundamentally, the issue in high dimensions is that a function's ``shape information'' --- of the type described in the preceding sections --- becomes more divorced from the value of its integral, making it more difficult to test the notion of ``sufficiently Gaussian shape to justify the LA''. There are a few different possible causes for this. The first is a well-known ``curse of dimensionality'' affecting certain high-dimensional probability density functions: most of their mass is in the tails, far away from the high-density region directly surrounding the mode \citep[e.g.][]{carpenter2017b, Betancourt2018}. Essentially, this happens because the neighbourhood around the mode is of a much smaller (Lebesgue) volume than the region encompassing the tails, so that most of the mass contributing to the integral is in a ``shell'' where the \textit{product} of density and volume is high [ibid.]. For instance, if $\mathbf{X}$ is a $d$-dimensional standard normal random variable, the \textit{Gaussian annulus theorem} \citep[][Theorem 2.9]{Blum2020} states that, with high probability, $\mathbf{X}$ will be in a spherical shell of width $\mathcal{O}(1)$ and distance $\mathcal{O}\left(\sqrt{d}\right)$ from the origin.

\begin{figure}
\centering\includegraphics[width=\textwidth, trim = {0 0.5cm 1.2cm 1cm}, clip]{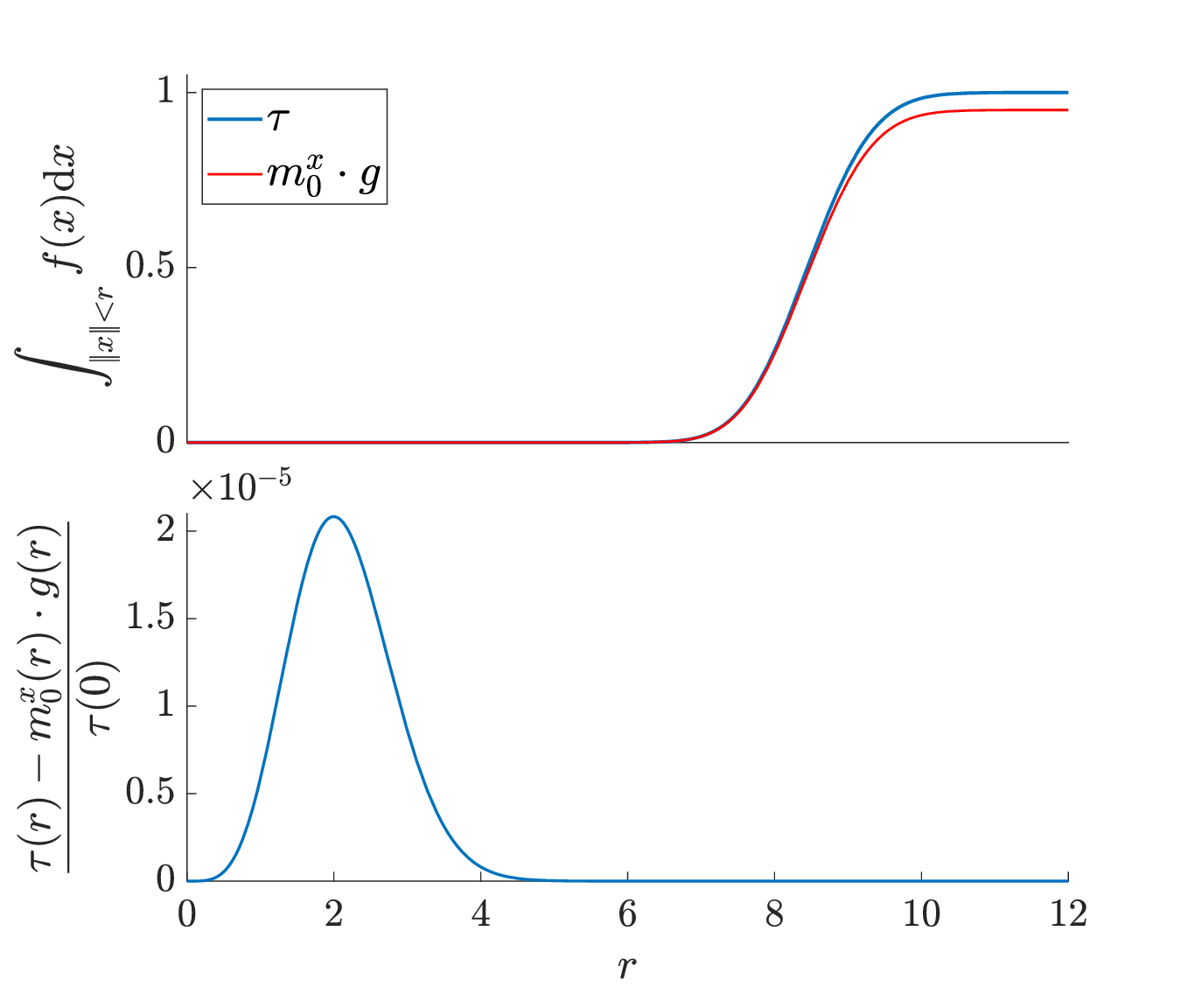}
\caption{Top: the amount of mass enclosed by $\tau$ = $\tau_{25921, 72}$ and its Gaussian approximation over a ball of radius $r$ centered at the origin. Bottom: the difference between $\tau$ and its Gaussian approximation at a distance of radius $r$ from the origin, normalized by the value of $\tau$ at the origin.}\label{fig:curse_dim}
\end{figure}

This poses an unfortunate challenge for the diagnostic: when the integrand $f$ is a high-dimensional density, its shape is easiest to visually assess around the mode where its values are relatively large, but its integral (and its LA, which is the integral of the Gaussian approximation to $f$) may be determined farther away where $f$ is much smaller. For example, consider the case $d = 72$ (the dimensionality of the real-data examples in Section \ref{sec:cod}), for which (as explained in Section \ref{sec:cal}) we take the calibration function $\tau$ to be a multivariate $t$ density with $\nu_{72} = 25921$ degrees of freedom because $L\left(\tau_{25921, 72}\right) = 0.95$. The top plot of \autoref{fig:curse_dim} shows the integral of this density --- and that of its Gaussian approximation ($m^x_0 \cdot g$, in the notation of Section \ref{sec:pn}) --- over the 72-dimensional ball $\{\mathbf{x}: \lVert \mathbf{x} \rVert < r\}$ as the radius $r$ varies. Observe that both $\tau$ and its Gaussian approximation have most of their mass between distances 7--10 from the origin. Furthermore, the difference between the integrals does not start to become apparent until the radius of integration is at least 8 (note that, as $r \to \infty$, the integrals of $\tau$ and its Gaussian approximation converge to 1 and the LA, respectively). This affirms the idea that most of the important information about the integral (in particular, its closeness to the LA) is quite far from the mode, in a region that authors such as \citet{Betancourt2018} call the \textit{typical set}. In contrast, the region of maximal \textit{shape difference} between $\tau$ and its Gaussian approximation occurs much closer to the origin, where there is almost no mass. This can be seen in the bottom plot of \autoref{fig:curse_dim}, which shows that $\tau$ differs most from its Gaussian approximation at a distance of approximately 2 from the origin. Even there, the largest difference between them is only about 0.002\% of $\tau$'s value at the mode. Further out in the aforementioned ``typical set'', the two functions are visually indistinguishable.

There is another interesting point to be made here about the high-dimensional diagnostic. It was stated in Section \ref{sec:cal} that $\nu_d$, the degrees of freedom for the calibration function in $d$ dimensions, would depend on $d$ itself. Indeed, the Laplace approximation for a multivariate $t$ density  (equation~\eqref{eq:mult_t_lap}) is decreasing in $d$ for fixed $\nu$. Thus, if $\nu_d$ is defined, as previously suggested, to be the smallest integer such that $L(\tau_{\nu_d, d}) \geq 0.95$, then $\nu_d$ is necessarily an increasing function of $d$. Put another way, in higher dimensions a $t$ density must be closer in shape to a Gaussian for its LA to be within 5\% of the true integral value. Indeed, using this definition of $\nu_d$ in 72 dimensions resulted in the extremely high value $\nu_{72} = 25921$. The difference between the resulting $t$ density and its Gaussian approximation is small enough to be virtually invisible, but because this difference is compounded over a (typical) set of extremely high volume, it results in a sizable difference between integrals.

In light of these ideas, our first suggested design for a high-dimensional diagnostic uses a preliminary grid $\mathbf{S}^* = \{\mathbf{0}\} \cup \{\pm\sqrt{d}\mathbf{e}_i: i \in \{1, \dots, d\}\}$, where $\mathbf{0}$ denotes the origin and $\mathbf{e}_i$ once again denotes the $i^\mathrm{th}$ standard basis vector of $\mathbb{R}^d$. This will result in $2d + 1$ interrogation points: one at the mode, and two at distances of $\mathcal{O}\left(\sqrt{d}\right)$ away from it along each ``principal axis'' (see Section \ref{sec:points}). Per the discussion above, if a function $f$ is assumed \textit{a priori} to have similar shape to a Gaussian density, then it is reasonable to expect this type of design to provide the most pertinent information about its integral. As described by \citet{Sarkka2015}, this choice of $\mathbf{S}^*$ in BQ creates a connection with \textit{sigma-point methods}, in which such grids are used to estimate integrals for filtering and smoothing in nonlinear SSM's \citep[e.g.][]{Julier1996}. In particular, aside from the inclusion of the origin this choice of $\mathbf{S}^*$ is identical to the point set used in the \textit{cubature Kalman filter} (CKF) of \citet{Ienkaran2009}.

With this preliminary grid in $d = 72$ dimensions, we use $\tau_{25921, 72}$ as our calibration function and once again take the hyperparameter $\gamma$ as in equation~\eqref{eq:gamma}, resulting in a value of $\gamma = 1.2248$. As alluded to above, here $\lambda$ cannot be selected to visually ensure that Condition (2a) is met as in the low-dimensional experiments of Section \ref{sec:2d}. Because the differences between the calibration function and its Gaussian approximation are so small at the chosen interrogation points, adjusting $\lambda$ does not produce any visible change in the difference $m^x_1 \cdot g - \tau_{25921, 72}$ (not shown). Thus, we must rely on the weaker Condition 2(b): selecting $\lambda$ to produce a reasonable posterior integral estimate $m_1$. We found $\lambda = 3.7$ to be a good choice for this, giving a posterior integral mean of $m_1 = 0.998$. Finally, $\alpha = 0.1565$ is once again chosen to ensure that the calibration curve's LA (equal to 0.95) is on the boundary of the rejection region. Note that, although we were unable to use shape information as directly as we did in the low-dimensional experiments, the diagnostic's rejection criterion still depends solely on the ``correction term'' (equation~\eqref{eq:correction}), itself a measure of deviation between a function and its Gaussian approximation. It could be said that the high-dimensional diagnostic, as it is configured here, determines whether a function is sufficiently Gaussian \textit{in the tails} to justify the LA.

\subsection{Example: North Sea cod modelling} \label{sec:cod}
This section returns to the SSM discussed in Sections \ref{sec:laplace_intro} -- \ref{sec:framework}. Recall that, given observed data $\mathbf{y}$, such a model can be fit by maximizing the Laplace-approximated marginal likelihood (integrating over hidden states $\mathbf{x}$) with respect to parameters $\mathbf{\theta}$. These methods are increasingly common in fisheries science, where they are used for \textit{stock assessment}: to infer population dynamics for various species of fish given observations from surveys and commercial catches \citep{Aeberhard2018}. SSM's applied to stock assessment are often called \textit{state-space assessment models (SSAM's)} [ibid.]\ and serve as a natural context to test our diagnostic: although the LA is commonly used in practice for these models, if the joint likelihood (equation~\eqref{eq:SSM_joint_lik}) is not ``sufficiently Gaussian'' in shape, then the LA may not be a suitable proxy for the marginal likelihood( equation~\eqref{eq:SSM_marg_lik}) and the resulting inferences may be incorrect.

To investigate the performance of our diagnostic in this ``real-world'' setting, we use a dataset containing multiyear measurements of cod stocks in the North Sea and fit SSAM's to various subsets of this data following \citet{Aeberhard2018}. The observations $\mathbf{y}_t$ are taken on an annual basis over the span of several decades ($t = 1963, \dots, 2015$). Briefly, for a given year $t$, $\mathbf{y}_t$ is a vector comprising the amounts of cod of different ages observed during surveys and commercial catches conducted that year. The hidden state $\mathbf{x}_t$ contains, for each age group, the ``true'' abundance and fishing mortality rate for cod in that age group in year $t$. Finally, $\mathbf{\theta}$ represents a variety of ``global'' parameters such as scaling factors and variances. The SSAM used here \citep[see][and references therein]{Nielsen2014} is highly nonlinear, with complex dependencies between the age-specific components of $\mathbf{x}_t$ and $\mathbf{x}_{t-1}$. For the sake of brevity further details are omitted here, but they are available in the appendix of \citet{Aeberhard2018}. All models were fit using the \texttt{stockassessment} R package \citep{Nielsen2014, Berg2016, R}, which is in turn built on the \texttt{TMB} package \citep{Kristensen2016}.
%\footnotetext{Note that the dimensionality of $\mathbf{y}_t$ is not constant with $t$, as the time ranges of the commercial catches and surveys only partially overlap. However, ``missing observations'' are not a problem for either model fitting or the diagnostic.}

Two SSAM's are considered here, each corresponding to a different subset of the available data: one fit to the data collected from 1970 to 1975 (hereafter the ``1970 model''), and another to the data from 2005--2011 (the ``2005 model''). Since each hidden state $\mathbf{x}_t$ is of dimension 12, using these six-year ``windows'' results in a latent dimensionality of $d = 12 \times 6 = 72$ for each model: fairly modest (and computationally convenient) compared to the $636$ dimensions associated with the full dataset \citep{Aeberhard2018}, but still large enough that any non-LA approach to marginalizing the likelihood would be far from trivial.
%\footnotetext{We also found that, with smaller time windows, there was not sufficient data to guarantee model convergence. Even six-year windows besides the ones used here typically did not converge without careful selection of algorithm settings.}
As stated above, the Laplace-approximated marginal likelihood $L\left(p_{xy}\right)$ is maximized numerically w.r.t.\ $\mathbf{\theta}$, and ideally we would like to use our diagnostic at each step of this optimization to ensure it remains accurate throughout. For simplicity in these experiments, we only apply the diagnostic at the last optimization step, seeking to determine \textit{only for the final parameter values} $\hat{\mathbf{\theta}}$ whether $p_{xy}(\cdot, \mathbf{y} \mid \hat{\mathbf{\theta}})$ is ``Gaussian enough'' to justify the LA.

In order to assess the performance of the diagnostic, it is desirable to have some other estimate of the marginal likelihood $p_y(\mathbf{y} \mid \hat{\mathbf{\theta}})$ to serve as an approximate ``ground truth''. Since standard numerical integration is completely nonviable in 72 dimensions, we instead obtain such estimates via importance sampling \citep[e.g.][and references therein]{Geweke1989}. %Possibly an unnecessary citation - I intended it as a citation for the very concept of IS, and I didn't do that when I mentioned it earlier
For both models, samples were taken from a noncentral multivariate $t$ distribution with mean $\hat{\mathbf{x}}$, %= \hat{\mathbf{x}}\left(\hat{\mathbf{\theta}}\right)$,
scale matrix $-\mathbf{H}^{-1}$, %= -\mathbf{H}^{-1}\left(\hat{\mathbf{\theta}}\right)$
and 5 degrees of freedom \citep{Evans1995}. The joint likelihoods of both models appear to have light tails in $\mathbf{x}$ (see below), so this choice of importance distribution should mitigate the risk of infinite variance in theory \citep{Yuan2007, Tokdar2009}. However, because we can only assess the tail behaviour of the models in finitely many directions, we cannot rule out the possibility that, somewhere in the 72-dimensional space, they have a tail even heavier than that of a $t$ density. We conjecture that this is not the case, although the existence of such a tail could result in a sampler with infinite variance. A more pressing concern is that poor finite-sample performance can still occur even with theoretical guarantees. Nevertheless, importance sampling is not the main concern here --- it is intended only as a convenient, if somewhat informal, check on the LA diagnostic.

This diagnostic is not the only way to check the LA for a SSM --- the \texttt{checkConsistency} function in the \texttt{TMB} package \citep{Kristensen2016} provides another method\footnotemark. It is essentially a \textit{score test} \citep{Rao1948} for the Laplace-approximated marginal likelihood: by simulating many separate datasets $\mathbf{y}^* \sim p_y\left(\cdot \mid \hat{\mathbf{\theta}}\right)$ (which can be done by simulating $\mathbf{x}^* \sim p_x$, then $\mathbf{y}^* \sim p_{y \mid x}$), it constructs a test statistic to test the hypothesis $\mathbb{E}_y[ \nabla_{\mathbf{\theta}} \log L(p_{xy}) \big\rvert_{\hat{\mathbf{\theta}}}] = 0$, under which the statistic would be asymptotically $\chi^2$-distributed. Since the true marginal score function has mean zero, %do I need a citation here, or is this just a math fact?
a rejection of this hypothesis means that the LA is \textit{not} a suitable approximation for the marginal likelihood $p_y$. It will be useful to compare this method to our diagnostic, but it should be noted that there is a key conceptual difference between them. The \texttt{checkConsistency} methodology views $L\left(p_{xy}\right)$ and $p_y$ as \textit{functions of} $\mathbf{y}$; with this view, it seeks to determine whether the marginal likelihood is well approximated by the LA, and what effects this approximation could have on the bias of the estimated $\hat{\mathbf{\theta}}$. In contrast, our diagnostic is focused on shape of the joint likelihood $p_{xy}$ when viewed \textit{as a function of} $\mathbf{x}$: in particular, whether this shape warrants the use of the LA to fit the model \textit{for the observed} (i.e.\ \textit{fixed}) $\mathbf{y}$.

\footnotetext{Refer to the source code at \url{https://github.com/kaskr/adcomp/blob/master/TMB/R/checker.R} for further detail. Notes provided by Anders Nielsen in personal correspondence also helped to inform this discussion.}

\begin{figure}
\centering{1970 model}
\centering\includegraphics[width=\textwidth]{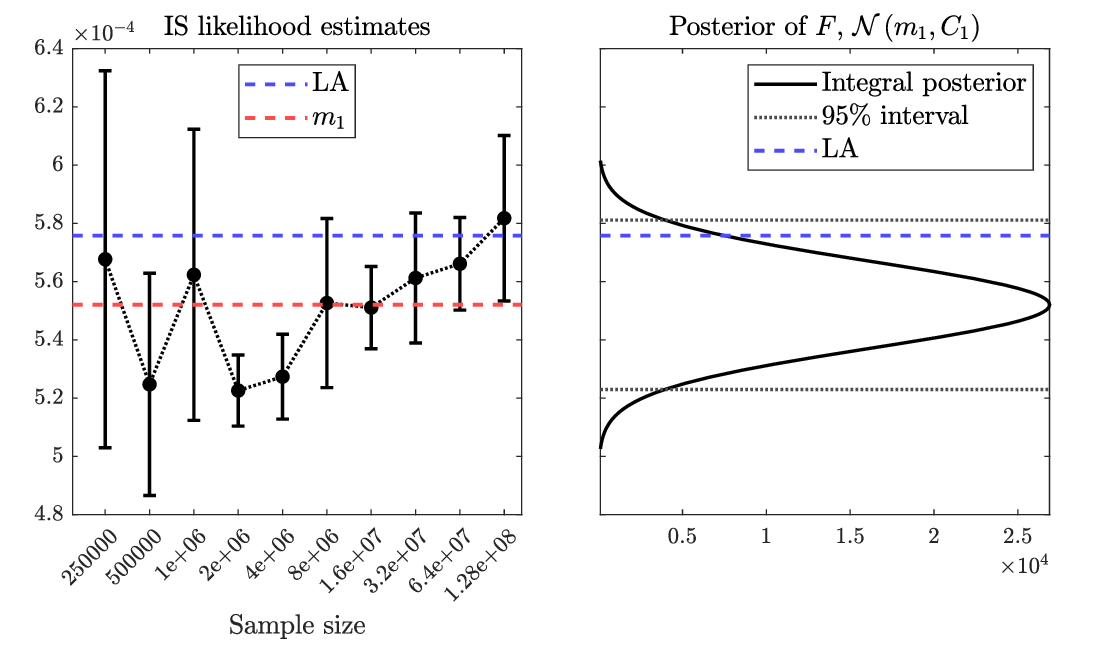}
% \centering\includegraphics[width=\textwidth, trim = {0.5cm 0 1cm 0}, clip]{1970_results.eps}
\caption{Results of the diagnostic applied to the 1970 SSAM. Left: IS estimates of $p_y\left(\mathbf{y} \mid \hat{\mathbf{\theta}}\right)$ at various sample sizes (black dots) with estimated 95\% confidence intervals (vertical line segments), with the Laplace approximation (blue dashed line) and the posterior integral mean (red dashed line) for reference. Right: the posterior distribution for the marginal likelihood, obtained from the diagnostic (rotated 90 degrees for ease of comparison with IS estimates).} \label{fig:1970_results}
\end{figure}

\autoref{fig:1970_results} shows results (from the diagnostic, as well as the aforementioned importance sampler with differing numbers of samples) for the 1970 model. For the importance samplers, 95\% confidence intervals were obtained with a Gaussian approximation, using the sample variance of the IS weights. The central limit theorem dictates that for a well-behaved importance sampler, the width of these intervals should be roughly $\mathcal{O}\left(S^{-1/2}\right)$, where $S$ is the number of samples. The left plot of \autoref{fig:1970_results} indicates that this may not be the case. Indeed, the score test of \citet{JanKoopman2009} rejected the hypothesis that these samplers had finite variance. These rejections are typically the result of a few large weights, which seemingly indicate that in a few directions the tails of $p_{xy}\left(\cdot, \mathbf{y} \mid \hat{\mathbf{\theta}}\right)$ are too heavy relative to those of the proposal density. However, further numerical evidence indicated that the tails of the squared joint likelihood were eventually dominated by its Gaussian approximation in those directions. In mathematical terms, at all sampled points $x \in \mathbb{R}^d$ for which the importance weights were large, it appeared that, for sufficiently large $r > 0$,
\begin{align}
\left[p_{xy}\left(\hat{\mathbf{x}} + rz, \mathbf{y} \mid \hat{\mathbf{\theta}}\right)\right]^2 = o\left(\phi\left(\hat{\mathbf{x}} + r\mathbf{z}\right)\right) \label{eq:bigO}
\end{align}
as functions of $r$, where $\phi$ is the Gaussian approximation to $p_{xy}\left(\cdot, \mathbf{y} \mid \hat{\mathbf{\theta}}\right)$ and $\mathbf{z}$ is a unit vector in the direction of $\mathbf{x} - \hat{\mathbf{x}}$. Since the ratio of a Gaussian density and a Student's $t$ density is certainly integrable over $\mathbb{R}^d$, this provides some limited indication that the integral defining the variance of the importance sampler \citep[e.g.][]{Evans1995} may indeed be finite after all. This is a very informal check on the validity of IS, and it does not guarantee finite-sample stability. However, their use as a heuristic reference against which to check the diagnostic does not seem unreasonable here.

Most of the importance samplers include the LA within their 95\% confidence intervals, suggesting it is not excessively far from the true marginal likelihood value. The fact that most of the IS estimates are below the LA suggests that the latter is perhaps a slight overestimate of the true value (i.e.\ that the tails of the joint likelihood, as a function of $x$, tend to be lighter than those of its Gaussian approximation). Our diagnostic produces a similar conclusion: the poste-rior integral mean is slightly lower than the LA, but not to a degree that warrants rejection. With respect to our notion of ``good-enough-ness-of-fit'', it seems that the LA is a reasonable approximation to the marginal likelihood for this model, at least for the parameter values $\hat{\mathbf{\theta}}$.

\begin{figure}
\centering{1970 model}
\centering\includegraphics[width=0.9\textwidth, trim = {0 0 0 -0.5cm}, clip]{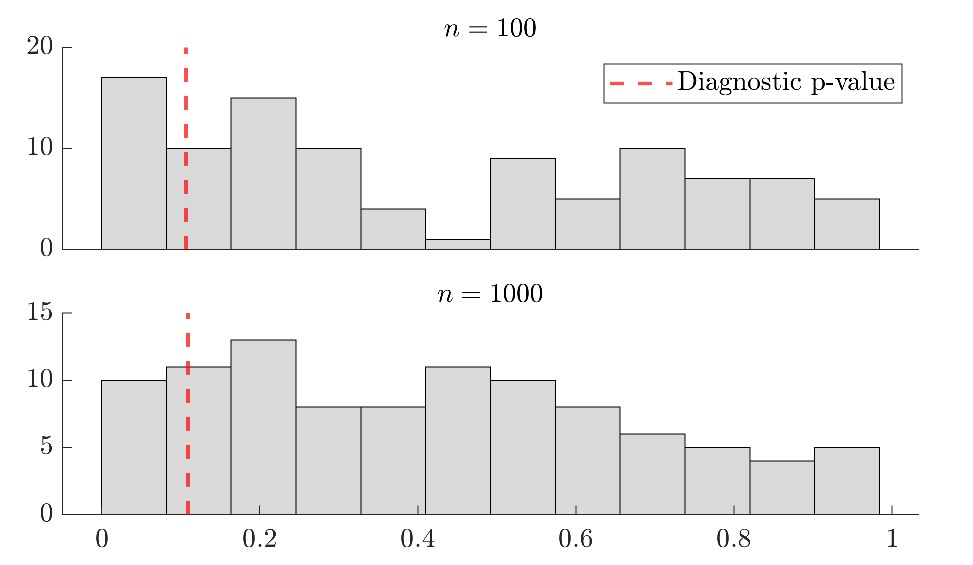}
% \centering\includegraphics[width=\textwidth]{1970_pvals.eps}
\caption{Histograms of p-values from repeated runs (100 runs each for simulation sizes $n = 100$ and $n = 1000$) of \texttt{checkConsistency} on the fitted 1970 SSAM. The ``p-value'' given by the diagnostic is shown as a dashed red line on each histogram.} \label{fig:1970_pvals}
\end{figure}

Since the diagnostic is based on a Gaussian ``confidence interval'' for the integral (see Section \ref{sec:pn}), its behaviour can be equivalently described in terms of ``p-values'': recalling from equation~\eqref{eq:int_posterior} that the integral posterior is $F \mid \mathbf{r}(\mathbf{S}) \sim \mathcal{N}\left(m_1, C_1\right)$, it is straightforward to show that the diagnostic rejects the LA iff
\begin{align}
2\left[1-\Phi\left(\frac{\left\lvert m_1 - L(f)\right\rvert}{\sqrt{C_1}}\right)\right] < 0.05, \nonumber
\end{align}
where $\Phi$ is the c.d.f.\ of a standard normal random variable, and the quantity on the left-hand side has a natural interpretation as a sort of ``p-value''. This facilitates some comparison between the diagnostic and the \texttt{checkConsistency} method. Recall that the latter simulates $n$ separate datasets to construct a test statistic that is asymptotically $\chi^2$-distributed when $\mathbb{E}_y[ \nabla_{\mathbf{\theta}} \log L(p_{xy}) \big\rvert_{\hat{\mathbf{\theta}}}] = 0$. This test statistic induces a p-value; if this is below some threshold (say, 0.05), we reject the hypothesis that the marginal likelihood and the LA are the same (as functions of $\mathbf{y}$). In \autoref{fig:1970_pvals}, we have performed the \texttt{checkConsistency} test 100 times each for two simulation sizes ($n = 100$ and $n = 1000$) in order to see how the p-value distribution changes with the number of simulated datasets and how it relates to the p-value of the diagnostic. If the null hypothesis of \texttt{checkConsistency} is true (i.e.\ the LA is the true marginal likelihood), then the p-value of the corresponding test should be uniformly distributed over $(0,1)$. Although the histograms in \autoref{fig:1970_pvals} show some deviation from uniformity, it is not severe. The p-value associated with the diagnostic is just above 0.1, consistent with non-rejection of the LA (see \autoref{fig:1970_results}). It is interesting to see from \autoref{fig:1970_pvals} that the diagnostic and \texttt{checkConsistency} seem to lead to similar conclusions --- that the LA may deviate slightly from the true marginal likelihood, but not to a problematic extent --- despite the fundamental difference in the questions addressed by each method.

\begin{figure}
\centering{2005 model}
\centering\includegraphics[width=\textwidth]{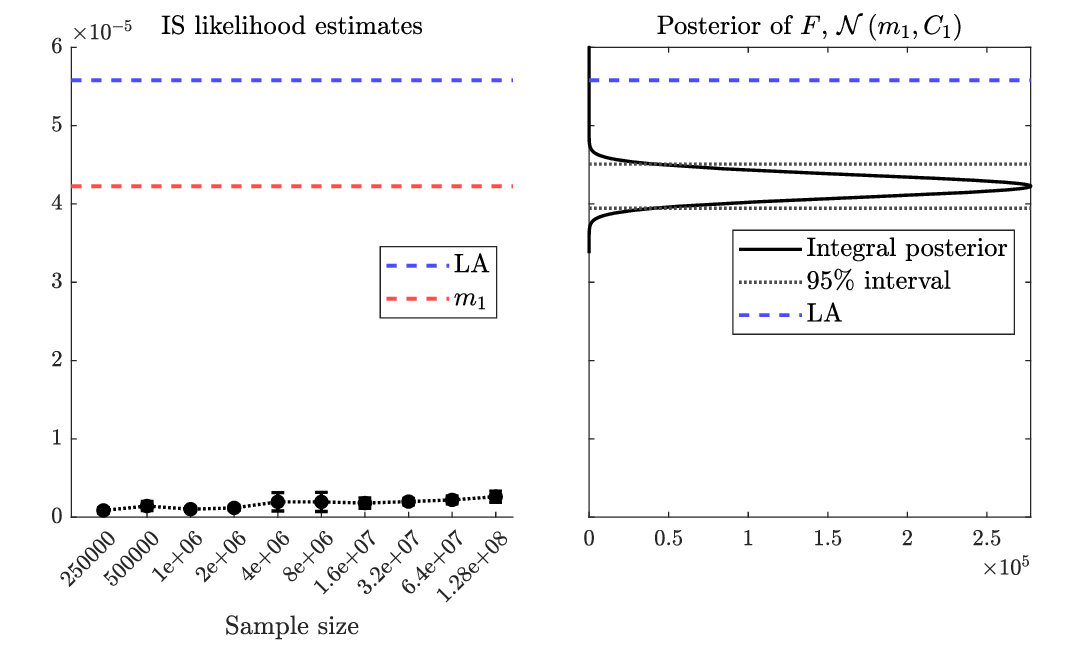}
\caption{Results of the diagnostic applied to the 2005 SSAM. Left: IS estimates of $p_y\left(\mathbf{y} \mid \hat{\mathbf{\theta}}\right)$ at various sample sizes (black dots) with estimated 95\% confidence intervals (vertical line segments), with the Laplace approximation (blue dashed line) and the posterior integral mean (red dashed line) for reference. Right: the posterior distribution for the marginal likelihood, obtained from the diagnostic (rotated 90 degrees for ease of comparison with IS estimates).} \label{fig:2005_results}
\end{figure}

\begin{figure}
\centering{2005 model}
\centering\includegraphics[width=0.9\textwidth, trim = {0 0 0 -0.5cm}, clip]{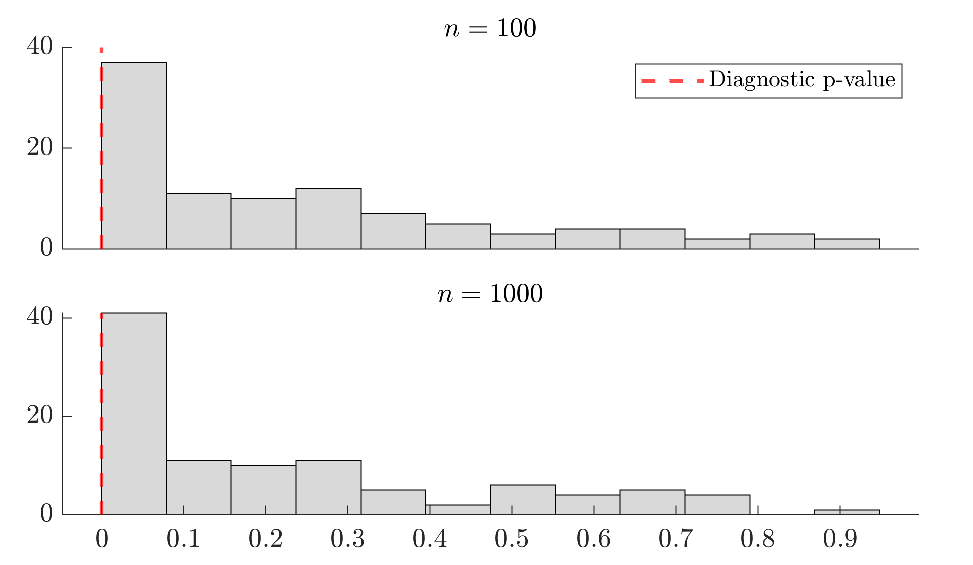}
\caption{Histograms of p-values from repeated runs (100 runs each for simulation sizes $n = 100$ and $n = 1000$) of \texttt{checkConsistency} on the fitted 2005 SSAM. The ``p-value'' given by the diagnostic is shown as a dashed red line on each histogram.} \label{fig:2005_pvals}
\end{figure}

The results are markedly different for the 2005 model, as shown in \autoref{fig:2005_results}. IS stability considerations apply here as they did for the 1970 model: Koopman et al.'s score test \citep{JanKoopman2009} rejected the hypothesis of finite variance for the largest sample sizes, but expression~\eqref{eq:bigO} appeared to hold in the directions of all the largest weights, potentially indicating a finite (but quite large) variance. All IS estimates are far lower than the LA, suggesting that the joint likelihood is, for the most part, substantially lighter-tailed than its Gaussian approximation. Accordingly, the diagnostic strongly rejects the LA, which is well above the upper bound of the posterior 95\% confidence interval. Note that there is still substantial disagreement between the diagnostic and the importance samplers as it pertains to estimation of the true marginal likelihood. Thus, the posterior integral mean from the diagnostic should not be taken as a high-quality estimate, but what is important is that both methods agree on rejection of the LA.

As before, we also conduct repeated runs of \texttt{checkConsistency} and compare the resulting p-value distributions to the one associated with the diagnostic. The latter is numerically indistinguishable from zero, and for both simulation sizes the p-value distribution is decidedly non-uniform. As was the case with the 1970 model, both methods appear to agree that the LA is an unsuitable approximation to the marginal likelihood, despite asking this question in different ways.

Differing philosophies notwithstanding, one clear advantage the diagnostic has over \texttt{checkConsistency} is computation time. Using the \texttt{checkConsistency} replications shown in Figures \ref{fig:1970_pvals} and \ref{fig:2005_pvals}, as well as 100 repeated computations of the diagnostic itself, \autoref{tab:time} shows median computation times --- along with median absolute deviations --- for each method applied to each model. All computations were performed on a computer with 64 GB of RAM and eight Intel i7-6400K 4GHz CPU cores. Note that the time cost for the diagnostic includes the evaluation of function interrogations, the eigendecomposition of the Hessian, and the calculation of all the necessary kernel terms for BQ (the latter step was sped up substantially using the methods of \citet{Karvonen2018}, as explained in Section \ref{sec:highgrid}). It is also interesting to note the differences in computational times between models: across all methods, the times for the 2005 model are longer than those for the 1970 model. Presumably, this is because of the ``inner'' numerical optimization \citep{Kristensen2016} used to calculate the mode $\hat{\mathbf{x}} = \hat{\mathbf{x}}\left(\mathbf{y}, \hat{\mathbf{\theta}}\right)$, which may require more iterations for the 2005 model than the 1970 model due to differences in their respective joint likelihoods. This would also explain why the difference is so much more pronounced for the \texttt{checkConsistency} runs, which require repeated (and possibly even more demanding) inner optimizations to find $\hat{\mathbf{x}} = \hat{\mathbf{x}}\left(\mathbf{y^*}, \hat{\mathbf{\theta}}\right)$ for each simulated dataset $\mathbf{y^*}$. In any case, the diagnostic is by far the fastest method of assessing the LA\footnotemark.
\footnotetext{IS computation times are not shown, as these were not replicated. However, they behaved largely as expected: computation times were roughly linear in the number of samples, and universally longer than those for the diagnostic.}

\begin{table}
\centering
\begin{tabular}{r|cc}
\toprule
Time (seconds)                                         & 1970 model        & 2005 model         \\
\midrule
\texttt{checkConsistency}, $n = 100$  & $2.511 \pm 0.035$  & $7.367 \pm 0.136$  \\
\texttt{checkConsistency}, $n = 1000$ & $25.115 \pm 0.152$ & $73.584 \pm 0.489$ \\
\bottomrule
Diagnostic                                             & $0.009 \pm 0.007$  & $0.012 \pm 0.0003$
\end{tabular}
\caption{Table showing median computation times (along with median absolute deviations) of each method, applied to each model.} \label{tab:time}
\end{table}

\subsection{Higher-order interrogation grids} \label{sec:highgrid}
The interrogation grids used thus far have been quite simple, consisting of $\mathcal{O}(d)$ preliminary points placed along the axes of $\mathbb{R}^d$ in a $d$-dimensional ``cross'' shape. As noted in Section \ref{sec:highdim}, there is precedent in the literature for the use of such simple grids \citep{Sarkka2015, Ienkaran2009}. They seem to be a reasonable choice here as well, allowing us to calibrate the diagnostic in such a way that appropriate results are obtained for a variety of ``toy'' and real-world examples. However, one potential drawback of such grids is that they only allow the diagnostic to use information about a function's shape along its ``principal axes'' (see Section \ref{sec:points}). If this is not indicative of the function's behaviour in the rest of the domain, it is conceivable that the diagnostic could produce misleading results. For instance, consider the $d$-dimensional function
\begin{align}
f_{\nu,d}(\mathbf{x}) = \prod_{i=1}^d\frac{\Gamma\left(\frac{\nu+d}{2}\right)}{\Gamma\left(\frac{\nu}{2}\right)\sqrt{\nu\pi}}\left(1+\frac{x_i^2}{\nu}\right)^{-\frac{\nu+d}{2}}. \label{eq:ind_t}
\end{align}
Like the multivariate $t$ density (equation~\eqref{eq:mult_t}), this function has a mode at the origin. The functions are equal there, as are the Hessians of their logarithms. Furthermore, they are equal along the axes of $\mathbb{R}^d$. Thus, their LA's are the same, and the diagnostic would give the same results for both functions using any of the ``cross-shaped'' interrogation grids considered above. However, the functions differ on the rest of their domain, and their integrals are different as a result. Whereas the integral of $\tau_{\nu,d}$ over $\mathbb{R}^d$ is equal to 1 for all $\left(\nu, d\right)$, the integral of $f_{\nu, d}$ is
\begin{align}
\frac{\Gamma\left(\frac{\nu + d - 1}{2}\right)^d}{\Gamma\left(\frac{\nu}{2}\right)\Gamma\left(\frac{\nu + d}{2}\right)^{d-1}}. \nonumber
\end{align}
In particular, for $d = 72,\ \nu = \nu_{72} = 25921$ (the values used to calibrate the 72-dimensional diagnostic at the beginning of this section), $\int_{\mathbb{R}^{72}} f_{25921, 72}(\mathbf{x}) \mathrm{d}\mathbf{x} = 0.952$. Thus the integral of $f_{25921, 72}$ is quite a bit closer to the LA (0.95) than that of the calibration function $\tau_{25921, 72}$, but the diagnostic calibrated with a ``cross-shaped'' grid will treat both of them identically, so that the LA is on the boundary of the rejection region for each function. One could argue that this is undesirable: the values of $f_{25921, 72}$ ``off the axes'' are lower (and therefore, closer to the Gaussian approximation) than those of the calibration function, causing its integral to be closer to the LA, so perhaps the diagnostic should produce a more definitive non-rejection for this function. For this to be possible, we must be able to capture the differences between $f_{\nu,d}$ and $\tau_{\nu,d}$, for which a \textit{higher-order} interrogation grid is required.

A grid of ``order'' $q$ is one whose size scales as $\mathcal{O}\left(d^q\right)$ for some fixed power $q > 1$ (the grids used throughout the manuscript thus far had $q = 1$). In order to use such grids without an excessive increase in computation time (which would defeat the purpose of the diagnostic), we use \textit{fully symmetric kernel quadrature} (FSKQ), as detailed by \citet{Karvonen2018}. Briefly, because the squared exponential kernel is isotropic, using fully symmetric preliminary grids (as described in Section \ref{sec:points}) reduces the number of \textit{unique} quadrature weights that need to be calculated, allowing for significant algebraic and computational simplifications in BQ.

\begin{figure}
\centering\includegraphics[width=0.7\textwidth, trim = {1.5cm 0 1.5cm 0}, clip]{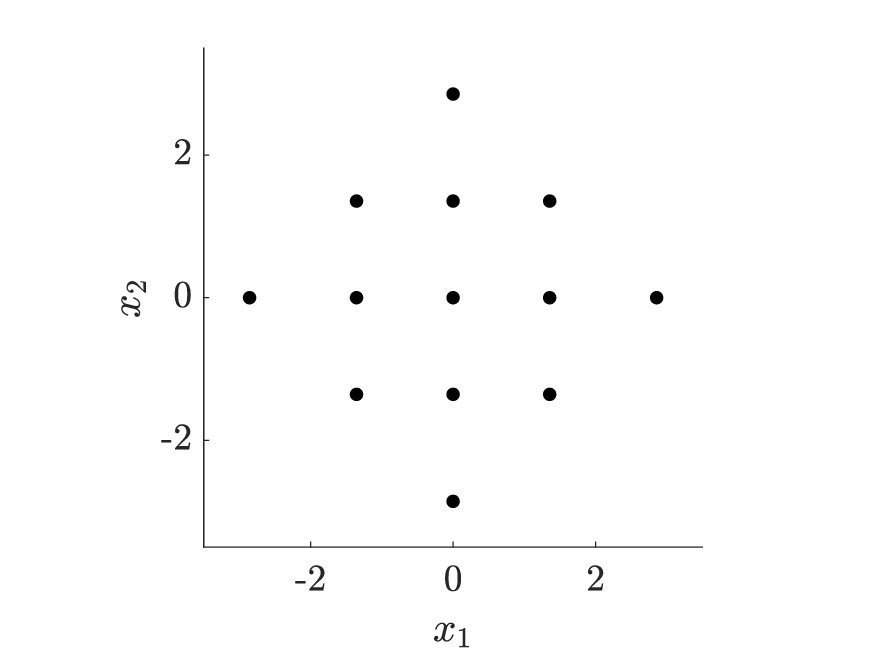}
% \centering\includegraphics[width=0.7\textwidth]{gauss_hermite.eps}
\caption{A sparse Gauss-Hermite quadrature grid of order 2 in $d = 2$ dimensions.} \label{fig:GH}
\end{figure}

Here, we conduct a few experiments with higher-order grids, showing difficulties associated with their use. We recalibrate the 72-dimensional diagnostic using a \textit{sparse Gauss-Hermite grid of order 2} --- the two-dimensional version of which is shown in \autoref{fig:GH} --- as the preliminary grid. Following \citet{Karvonen2018}, we remove the origin, as its quadrature weight tends to be a large negative value for most hyperparameter combinations. Furthermore, because a function is always equal to its Gaussian approximation at the mode, the origin does not actually contribute to the diagnostic beyond its effect on the inverted Gram matrix. We also multiply each point in the Gauss-Hermite grid by 3.6, thereby ensuring that they are far enough away from the origin to cover the ``typical set'' discussed in Section \ref{sec:highdim}. The final preliminary grid in $72$ dimensions is of size $n = 10512$, and as with the  original ``cross-shaped'' preliminary grid (which, for reference, contained $n = 145$ points) we calibrate the diagnostic using the $t$ density $\tau_{25921, 72}$ and taking the hyperparameter $\gamma = 1.2248$. As before, it is not possible to calibrate with respect to Condition (2a) from Section \ref{sec:cal}. Here, this is because of the size of the grid: the computational simplifications of FSKQ are only applicable to the integral of the GP, not to the GP posterior mean function (equation~\eqref{eq:mean}) itself. As such, the visual calibration of Section \ref{sec:2d} is not viable: even though we would only need to view a 2-dimensional slice of $m^x_1\cdot g - \tau_{25921, 72}$, every change to the hyperparameter $\lambda$ would still necessitate the recalculation and inversion of the $10512 \times 10512$ Gram matrix, which is too slow for minute visual adjustments. Instead, we once again calibrate with respect to Conditions (1) and (2b), resulting in hyperparameters $\left(\lambda, \alpha\right) = (3.7, 0.1349)$ and a posterior integral mean of $m_1 = 0.9945$ for the calibration function.

\begin{figure}
\centering{1970 model}
% \centering\includegraphics[width=\textwidth, trim = {0.5cm 0 1cm 0}, clip]{1970_GH.eps}
\centering\includegraphics[width=\textwidth]{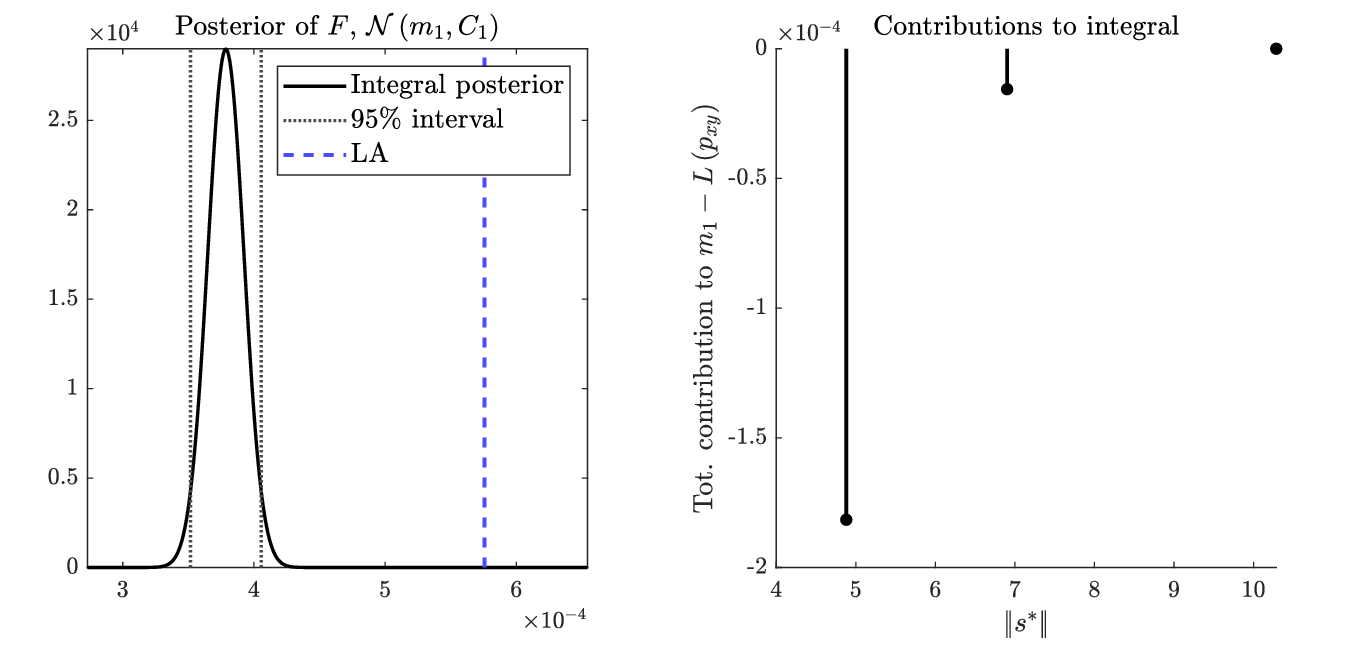}
\caption{Results of the diagnostic with a higher-order interrogation grid applied to the 1970 SSAM. Left: the posterior distribution for the marginal likelihood, obtained from the diagnostic. Right: the total mass contributions to the quadrature estimate made by interrogations as a function of the distance between the corresponding preliminary points and the origin.} \label{fig:1970_GH}
\end{figure}

Applying the new calibrated diagnostic with the larger preliminary grid to the SSAM's from Section \ref{sec:cod} reveals that the use of higher-order grids does not necessarily cause an improvement in the diagnostic's behaviour in practice --- indeed, the opposite may occur. The left plot of \autoref{fig:1970_GH} shows that, in contrast to the results in Section \ref{sec:cod}, this version of the diagnostic rejects the LA for the 1970 model. Initially, this may suggest that the tails of the joint likelihood are substantially lighter than those of its Gaussian approximation in directions besides its ``principal axes'', which would not have been observable using the smaller grid. However, this is at odds with the results of the importance samplers and \texttt{checkConsistency}, both of which suggested that the LA was \textit{not} very far from the true marginal likelihood and neither of which is constrained to the use of information on the principal axes of the joint likelihood. Furthermore, the right plot of \autoref{fig:1970_GH} reveals that the largest overall contribution to the lowered integral estimate comes from the interrogation points which are closest to the mode. This is despite the fact that there are only 144 such points in the Gauss-Hermite grid. In contrast, the points further from the origin --- of which there are 10368 --- collectively contribute a much smaller amount to the estimate. As discussed in Section \ref{sec:highdim}, the integral of a high-dimensional function is mainly determined by the behaviour of its tails; ideally this would be reflected when using a preliminary grid with most of its points far away from the origin. In light of these considerations, it seems reasonable to conclude that this version of the diagnostic is not providing accurate inference on the integral, or on the function shape information most pertinent to it.

\begin{figure}
\centering{2005 model}
% \centering\includegraphics[width=\textwidth, trim = {0.5cm 0 1cm 0}, clip]{2005_GH.eps}
\centering\includegraphics[width=\textwidth]{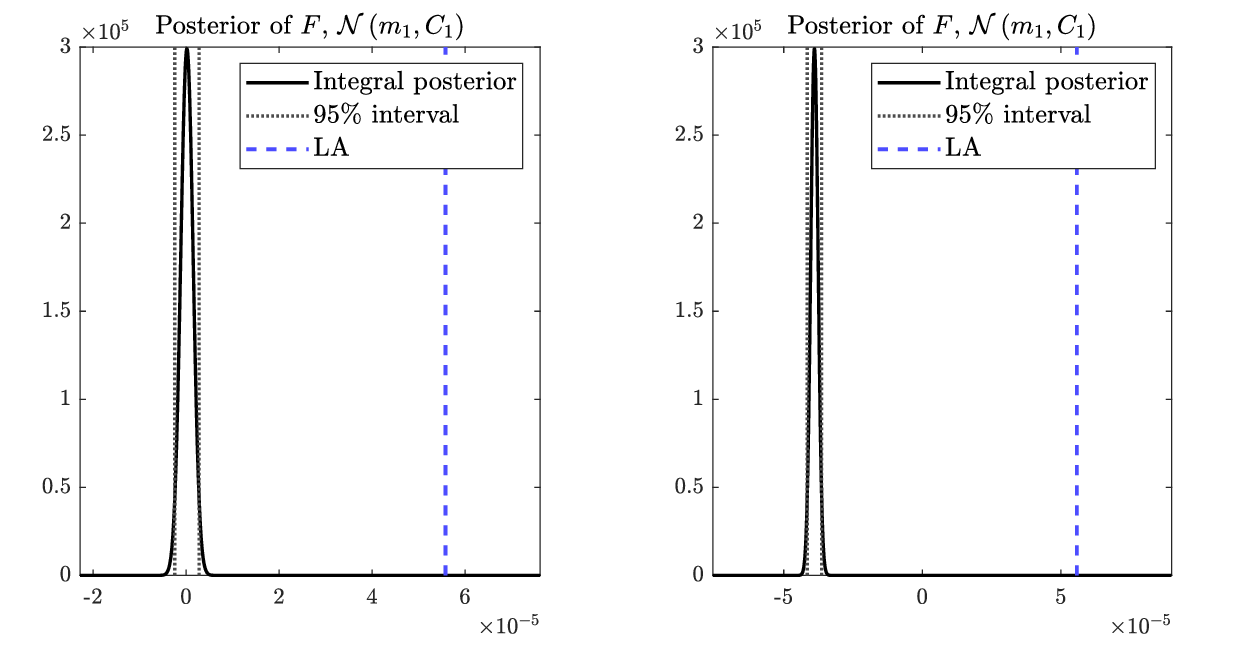}
\caption{Results of the diagnostic with a higher-order interrogation grid applied to the 2005 SSAM. Left: the posterior distribution for the marginal likelihood, obtained from the diagnostic. Right: the same, but with a single interrogation point having been removed.} \label{fig:2005_GH}
\end{figure}

The new diagnostic exhibits a different problem when applied to the 2005 model, as seen in \autoref{fig:2005_GH}. The left plot shows that the LA is once again definitively rejected, although the actual integral posterior differs quite noticeably from the one in \autoref{fig:2005_results}. However, as it turns out, there is one interrogation point $\mathbf{s}$ where the weighted difference $r(\mathbf{s}) - m_0^x(\mathbf{s})$ is far larger than it is for any of the other points. Removing this point from the grid, but keeping the hyperparameters fixed\footnotemark\footnotetext{Note that deleting the corresponding preliminary interrogation point did not produce a sizeable change in the diagnostic's behaviour when applied to the calibration function (not shown), despite not adjusting the hyperparameters for the altered grid. Thus, there is no concern about miscalibration here.}, results in a surprisingly large change in the posterior, shifting its mean from a small positive value to a larger negative value (which is nonsensical, given that the integral is a likelihood and must therefore be nonnegative). Although the diagnostic achieves its primary goal in both cases for this model --- namely, determining that the joint likelihood's shape (as a function of $\mathbf{x}$) is too non-Gaussian to justify the LA --- it is certainly undesirable for one interrogation point to have such a large impact. If this were allowed, a given function's LA could be rejected based solely on the inclusion or exclusion of a single point at which it deviates significantly from its Gaussian approximation, thereby rendering the diagnostic too sensitive to be useful for nontrivial high-dimensional applications (see the discussion at the beginning of Section \ref{sec:design}).

At first, the failure of the diagnostic with high-order interrogation grids seems illogical. Intuitively, one would expect more accurate quadrature with larger grids. Indeed, several convergence theorems in the BQ liteature suggest that the addition of more points should be an asset \citep[e.g][]{Briol2015, Karvonen2018, Briol2019}. However, these results tend to assume that the kernel and integrating measure are fixed. Here, we change both through our calibration of the hyperparameters, a necessary step in fulfilling the goals of the diagnostic. In this instance, asymptotics fail to guarantee the type of practical, finite-sample behaviour we require. Despite the potential shortcomings of lower-order grids, they seem to be a better choice in terms of ensuring a usable diagnostic, unless great care is taken with higher-order grids.

The computation times for the diagnostic with the higher-order grid are predictably higher than they were for the original diagnostic, although it is still much faster than \texttt{checkConsistency}. The median time was 0.4154 seconds for the 1970 model (MAD: 0.0105 seconds) and 0.5422 seconds for the 2005 model (MAD: 0.0104 seconds). Nevertheless, given the difficulties encountered above, the simpler CKF-style grid used in Section \ref{sec:cod} seems to be a better choice.

\section{Discussion}
In this manuscript, we have built on the work of \citet{Zhou2017} to develop a non-asymptotic diagnostic tool for assessing the viability of Laplace approximations to integrals. More specifically and accurately, the diagnostic assesses whether a function's shape is close enough to the Gaussian approximation that is used to motivate the LA. It does so using the method of Bayesian quadrature, but in multiple ways it is structured differently than a more ``conventional'' BQ application. Namely, we avoid design choices that would ensure accurate, low-uncertainty estimates for the integral of a specific function, opting instead for a ``one-size-fits-all'' approach: relatively simple interrogation grids intended to capture the most pertinent information about a function's behaviour, hyperparameters chosen heuristically using calibration functions, and a covariance structure that ensures the diagnostic is invariant to all properties of the integrand besides its shape. More broadly, the diagnostic is based on a notion of ``good-enough-ness-of-fit'' that stands in stark contrast to a more conventional, power-focused approach to statistical inference. Indeed, such an approach would render the diagnostic useless, causing it to prioritize the detection of \textit{any} deviation from Gaussian shape and likely producing rejections in almost all non-trivial applications.

As shown in this paper, challenges arise when using the diagnostic in high dimensions, although they are not insurmountable. Compared to the low-dimensional settings of Sections \ref{sec:cal}--\ref{sec:banana}, it is more difficult to make conclusions about a function's integral given limited information about its shape --- either because a high-dimensional function's mass tends to be far away from the regions with the most notable ``shape information'' (the curse of dimensionality), or because a single direction of non-Gaussian shape (which, intuitively, seems more likely to occur in high dimensions) can affect the diagnostic's behaviour to an unreasonable extent. Because of these challenges, more consideration must be given in high-dimensional spaces when choosing the preliminary interrogation grid and setting the hyperparameters, and the focus must be on the function's shape in its tail regions, assumed to correspond to its ``typical set''. If this is done carefully, the diagnostic can be calibrated to produce reasonable and useful results on real-world examples, as shown in Section \ref{sec:cod}.

Given SSAM's that had already been fit (producing parameter estimates $\hat{\mathbf{\theta}}$), we applied the diagnostic to their joint likelihoods $p_{xy}\left(\cdot, \mathbf{y} \mid \hat{\mathbf{\theta}}\right)$. While this served the purposes of this manuscript (namely, a proof-of-concept for the diagnostic itself), it ignores the fact that the parameter estimate itself depends on the use of Laplace approximations: specifically, that it is obtained by maximizing the LA $L\left(p_{xy}\left(\cdot, \mathbf{y} \mid \mathbf{\theta}\right)\right)$ with respect to $\mathbf{\theta}$. Given the low computational cost of the diagnostic, it would be desirable to fold it directly into a model-fitting workflow, checking at each iteration of numerical optimization whether or not the LA is justified, thereby indicating if other methods need to be invoked to correct any incurred bias in the estimated model parameters.

Despite the promising initial performance of the diagnostic, there are opportunities for future potential improvements. The difficulties of using higher-order grids encountered in Section \ref{sec:highgrid} should be further explored, as their resolution could result in improved diagnostic behaviour on a wider variety of functions. The methods of choosing interrogation points cited in the introduction of Section \ref{sec:design} may be a useful starting point to this end, but care must be taken to modify these methods in a way that preserves the quick, ``one-size-fits-all'' nature of the diagnostic. Another aspect of the diagnostic that remains unaddressed is the prior structure: specifically, that our use of a GP prior is \textit{technically} inappropriate given that most applications involve likelihoods, which are nonnegative. It is worth investigating other prior specifications proposed in the BQ literature \citep[e.g.][]{Gunter2014, Chai2019}, which preserve nonnegativity of the integrand at the expense of inducing a non-analytic distribution on the integral which must be approximated.

As a final note, we conjecture that the methods developed here may be more broadly applicable beyond the assessment of Laplace approximations. Indeed, a great deal of statistical methods are based on an assumption that some function is well approximated by a Gaussian shape, which is precisely the assumption that the diagnostic is designed to check. The general idea of using non-asymptotic methods to diagnose the use of asymptotic methods is one that warrants further consideration and study.

\section*{Acknowledgements}
Shaun McDonald wishes to thank the defense committee who reviewed the Ph.D.\ thesis in which this work originally appeared. We also wish to thank Richard Lockhart for providing insight on a suitable form for the covariance kernel.

\section*{Declarations}
This manuscript is largely an adaptation of the doctoral thesis of \citet{mcdonald2022}.
%\bmhead{Conflict of interest} The authors have no conflicts of interest to report.
%\bmhead{Funding} Both authors are supported by NSERC.

\bibliographystyle{plainnat}
\bibliography{laplace-approx-bib}

\end{document}